\documentclass{article}  

\usepackage{graphicx}

\usepackage{geometry}
\usepackage{txfonts}
\usepackage{amssymb}
\usepackage{amstext}
\usepackage{natbib}

\usepackage{hyperref}
\newcommand{\jgr}{J. Geophys. Res. }
\newcommand{\grl}{Geophys. Res. Lett. }
\newcommand{\icarus}{Icarus }
\newcommand{\aap}{Astron. Astrophys. }
\newcommand{\apj}{Astrophys. J. }

\newcommand{\ssr}{Space Sci. Rev. }

\newcommand{\nat}{Nature }
\newcommand{\science}{Science }

\newcommand{\jqsrt}{J. Quant. Spectrosc. Radiat. Transfer }

\newcommand{\natastron}{Nat. Astron. }

\newcommand{\natcommun}{Nat. Commun.} 
\newcommand{\jms}{J. Mol. Spectrosc.} 
\newcommand{\sciadv}{Sci. Adv.} 

\newcommand{\m}[1]{\,m$^{#1}$}

\newcommand{\dix}[1]{$\times10^{#1}$}
\newcommand{\fig}[1]{Fig.~\ref{#1}}
\newcommand{\figs}[2]{Figs.~\ref{#1} and \ref{#2}}

\newcommand{\degre}{\ensuremath{^\circ}}

\begin{document} 


\Large
\begin{center}\textbf{Observations of the temporal evolution of Saturn's stratosphere following the Great Storm of 2010-2011. II. Latitudinal distribution of CO and stratospheric winds}\end{center}\normalsize

\large\noindent T. Cavali\'e$^{1,2}$, R. Moreno$^{2}$, C. Lefour$^{1}$, B. Benmahi$^{3}$, T. Fouchet$^{2}$, E. Lellouch$^{2}$, \'E. Ducreux$^{4,5,6}$, M. Gurwell$^{7}$, F. Gueth$^{8}$, L. N. Fletcher$^{9}$, D. Bardet$^{10}$\normalsize\\
\vspace{0.2cm}

\noindent$^1$Laboratoire d'Astrophysique de Bordeaux, Univ. Bordeaux, CNRS, B18N, all\'ee Geoffroy Saint-Hilaire, 33615 Pessac, France (ORCID: 0000-0002-0649-1192, email: thibault.cavalie@u-bordeaux.fr)\\ 
$^2$LIRA -- Laboratoire d'Instrumentation et de Recherche en Astrophysique, Observatoire de Paris, Section de Meudon, 5, place Jules Janssen -- 92195 MEUDON Cedex\\
$^3$ Aix-Marseille Universit\'e, CNRS, CNES, Institut Origines, LAM, Marseille, France\\
$^4$ Universit\'e de Reims Champagne-Ardenne, CNRS, GSMA, Reims, France\\
$^5$ Planetary Atmospheres, Royal Belgian Institute for Space Aeronomy, Avenue Circulaire, 1180 Brussels, Belgium\\
$^6$ Institute of Life, Earth and Environment (ILEE), University of Namur (UNamur), 61 rue de Bruxelles, Namur, 5000, Belgium\\
$^7$ Center for Astrophysics | Harvard \& Smithsonian, 60 Garden Street, Cambridge, MA 02138, USA\\
$^8$ Institut de Radioastronomie Millim\'etrique (IRAM), 38406, Saint-Martin-d'H\`eres, France\\
$^9$ School of Physics and Astronomy, University of Leicester, Leicester, UK\\
$^10$ Laboratoire de M\'et\'eorologie Dynamique,  Ecole Normale Sup\'erieure, 24 rue Lhomond, 75231 Paris France\\

\vspace{0.2cm}
\noindent\textbf{Received:} 29 August 2025\\
\noindent\textbf{Accepted:} 23 January 2026\\
\vspace{0.2cm}

\noindent\textbf{DOI:} 10.1051/0004-6361/202557032 \\
\vspace{0.5cm}

\section*{Abstract}
   Saturn's Great Storm of 2010-2011 has produced two stratospheric hot spots, the ``beacons,'' that eventually merged to produce a gigantic one in April and May 2011. This beacon perturbed stratospheric temperatures, hydrocarbon, and water abundances for several years. We aim to assess whether the beacon induced any perturbation in another oxygen species, namely CO. A second goal is to measure how the vortex perturbed the stratospheric wind regime. We conducted interferometric observations of Saturn in the submillimeter range with SMA and ALMA to spatially resolve the CO (J=3-2) and (J=2-1) emissions, respectively. We used a previously determined CO vertical profile as a template, to search for (i) the meridional distribution of CO and (ii) variations of the CO abundance associated with the storm. The high spatial and spectral resolutions of the ALMA observations enabled us to retrieve the winds from the Doppler shifts induced by the winds on the lines. Despite limitations resulting from the removal of baseline ripples, we find a relatively constant meridional distribution of CO. The average CO mole fraction implied by the adopted and rescaled 220-year-old-comet-impact vertical profile is (1.7$\pm$0.7)\dix{-7} at 0.3\,mbar, i.e., where the contribution functions peak. We also find that the CO abundance has not been noticeably altered in the beacon. The winds measured at 1\,mbar show striking differences with those measured in 2018, after the demise of the beacon. We find the signature of the vortex as an anticyclonic feature. The equatorial prograde jet is 100 to 200 m.s$^{-1}$ slower, and broader in latitude, than in quiescent conditions. We also detect several prograde jets in the southern hemisphere. Finally, we detect a retrograde jet at 74$^\circ$N which could be a polar jet caused by the interaction of the Saturn magnetosphere with its atmosphere. With Saturn's equinox season approaching, new wind measurements would enable the findings presented in this paper to be confirmed by probing the two hemispheres equally and searching for a southern retrograde polar jet.

\section{Introduction}
Saturn's Great Storm of 2010-2011 \citep{Sanchez-Lavega2011,Fischer2011,Fletcher2011} was remarkable in several aspects. It was the longest global storm ever observed in Saturn's atmosphere \citep{Sanchez-Lavega2018}, with a total duration of more than 6 months at cloud level. The initial perturbation was observed in December 2010 at 40$^\circ$N planetocentric and spread over all longitudes within weeks to eventually encircle the whole planet \citep{Sanchez-Lavega2011,Sayanagi2013}. Also compelling is the fact that the storm produced two stratospheric hot spots \citep{Fletcher2011}, nicknamed ``beacons,'' due to their strong thermal infrared emission and to the rapid rotation of the planet. While one beacon was positioned on top of the initial tropospheric perturbation (i.e., at higher altitudes, but at similar latitudes and longitudes), the other was located on top of the storm's cloudy tail. When the storm head and tail merged at the cloud level, the two stratospheric beacons merged as well, leading to a gigantic hot spot spanning $\sim$40$^\circ$ in latitude and $\sim$100$^\circ$ in longitude, in which the temperature contrast reached a peak of more than 80\,K above quiescent conditions at the mbar level \citep{Fletcher2012b}. 

The merged beacon lasted approximately 3 years \citep{Lefour2025}, with its lifetime expanding way beyond the lifetime of the tropospheric perturbation. And while there is growing evidence that the troposphere below the clouds had its composition altered in the long term \citep{Li2023b}, the Saturn stratosphere also underwent significant changes in minor species abundances. Several hydrocarbons, such as ethane, acetylene, and ethylene, have seen their abundances increased within the beacon \citep{Fletcher2012b,Hesman2012}. The chemistry coupled with the increase in temperature alone cannot account for these departures from the normal conditions \citep{Moses2015,Cavalie2015}, and downward vertical winds interior to the vortex had to be invoked \citep{Moses2015}. In addition, the huge temperature increase monitored inside the vortex \citep{Fletcher2012b,Fletcher2018b} led to the sublimation of the stratospheric water haze \citep{Lefour2025} resulting from the condensation of externally sourced water \citep{Cavalie2019} supplied by Enceladus and its water torus \citep{Hartogh2011a,Cassidy2010}. Together with downwelling winds, this has led to an increase by a factor of $\sim$7.5 in the water column density in the beacon over several months.

Carbon monoxide (CO) has a dual origin in Saturn. It has an internal source \citep{Fouchet2017} that is caused by the deep thermochemistry of water and methane \citep{Cavalie2024}. In the stratosphere, an external source was identified \citep{Cavalie2009} and was attributed to an ancient large comet impact \citep{Cavalie2010} which seems to be common in the Solar System \citep{Lellouch1995,Lellouch2005,Cavalie2014}. More recently, the Cassini Ion and Neutral Mass Spectrometer (INMS) instrument, during the final and proximal orbits of the mission, revealed a strong flux, sharply localized in the equatorial plane, falling from the innermost D ring onto Saturn's upper atmosphere \citep{Waite2018,Perry2018}. This source contains mass-28 material, which could in part be composed of CO. It is tentatively thought to be linked to the breakup in 2015 of a small moon in the D68 ringlet \citep{Hedman2016}. While the storm was probably not powerful enough to lift CO-containing material from the troposphere to the stratosphere, as shown by the lack of ammonia signatures in the stratosphere in storm data \citep{Sromovsky2013}, the vertical winds derived by \citet{Moses2015} should nonetheless have altered its stratospheric distribution locally, as has been the case for hydrocarbons. One of the goals of this paper is to assess whether CO underwent any changes in its stratospheric distribution, by mapping it with the Submillimeter Array (SMA) before the onset of the storm and with the Atacama Large Millimeter/submillimeter Array (ALMA) during the storm.

When observed in the submillimeter wavelength range, CO can also be used as a tracer for stratospheric winds in the 0.1-10\,mbar region \citep{Benmahi2022}. The retrieval of stratospheric winds from spectral line observations requires the measurement of Doppler shifts induced by the winds on the spectral lines for a set of positions on the planet. Heterodyne spectroscopy in the millimeter range with an interferometer offers the necessary spatial and spectral resolutions to do so. A second goal of this paper is then the direct measurement of the stratospheric winds in Saturn's stratosphere during the storm from the ALMA data.

In this paper, we present our pre-storm and storm mapping observations of CO in section \ref{sec:Observations}. We describe our models in section \ref{sec:Models}, and explain how we derived the CO meridional distribution and discuss any changes seen between the pre-storm and storm conditions in section \ref{sec:Meridional}. Then, we used these results to derive the stratospheric winds during the storm and discuss them in section \ref{sec:Winds}. We give our conclusions in section \ref{sec:Conclusions}.

\section{Observations}\label{sec:Observations}
  \subsection{March 2010 SMA observations}
  We targeted the emission of the CO (J=3-2) line at 345.796\,GHz from Saturn's atmosphere with the Submillimeter Array, as part of project 2009B-S037 in a single track on March 13, 2010. The planet had an equatorial angular size of 19.52'', and a sub-Earth latitude of 3.4$^\circ$N\footnote{All latitudes given in this paper are planetocentric latitudes.}. Due to the large scale of the Saturn disk, we used the inner six antennas of the subcompact configuration providing 15 baselines, with minimum and maximum distances of 6-m and 25.3-m, respectively. The weather was exceptionally good, with 225\,GHz zenith opacity consistently below 0.04 for the entire observation (with a mean opacity of 0.03), corresponding to a mean of atmospheric precipitable water vapor of 0.4\,mm. Based on the mean 225-GHz opacity, we estimated the atmospheric transmission at the CO (J=3-2) frequency to amount to 0.89.
  
  To calibrate the observations, the flux density standard used was Titan (the model corresponds to that described in \citealt{Butler2012}) along with Mars. Amplitude and phase gains were measured using observations of the bright blazar 3C273 (approximately 8.7\,Jy at the CO (J=3-2) frequency at the time of the observations). Passband calibration was determined using a combination of observations of Mars, Saturn, and 3C273, in a several step process (see below).  
  
  SMA operated double sideband (DSB) mixers, with sideband separation occurring after local oscillator (LO) downconversion through a standard phase switching technique. The total bandwidth covered was $\sim$4\,GHz in each sideband, comprised of forty eight 104\,MHz windows (spaced by 82\,MHz). Here we focus only on the seven upper sideband windows around the CO (J=3-2) line, i.e., a central 82\,MHz window, and $\pm$3 windows on each side for an analyzed bandwidth of 574\,MHz, and the corresponding lower side band windows, which utilized the same intermediate frequency (IF) signal path after the initial LO downconversion, and thus have common spectral response to two IF filtering steps with significant spectral characteristics in both amplitude and phase. The spectral resolution is 812.5\,kHz. During the observations, the LO frequency was shifted to track the Doppler-shifted position of the CO line, preventing smearing from the changing observatory-Saturn velocity. The final spectral cube has the frequencies scale as observed at the observatory and requires final correction by the observatory-planet relative velocity, which we obtain from JPL Horizons ephemerides.
  
  Passband calibration of the visibility data was challenging for two reasons. The first issue was that the flux density of Saturn (primary beam attenuated full disk $\sim$2700\,Jy), even on heavily resolved baselines, was significantly higher than even the brightest available standard calibration sources (typically bright blazars). From a signal-to-noise ratio (S/N) standpoint calibration, calibration observations would need to proceed for much longer than science observations. To ameliorate this, we observed not only the bright blazar 3C273 (flux density $\sim$8.7\,Jy), but also Mars (diameter 10.9'', full disk flux $\sim$1500\,Jy). The second issue was that, at the time the SMA signal path, especially for bright continuum sources, exhibited strong standing waves in the passband, due to a variety of factors including impedance mismatches and internal reflections, and incomplete autoscaling of input power to optimize system performance. These standing waves change due to changes in the input system temperature as well as mechanical changes.
  
\begin{figure}[!ht]
    \centering
    \includegraphics[width=0.7\textwidth]{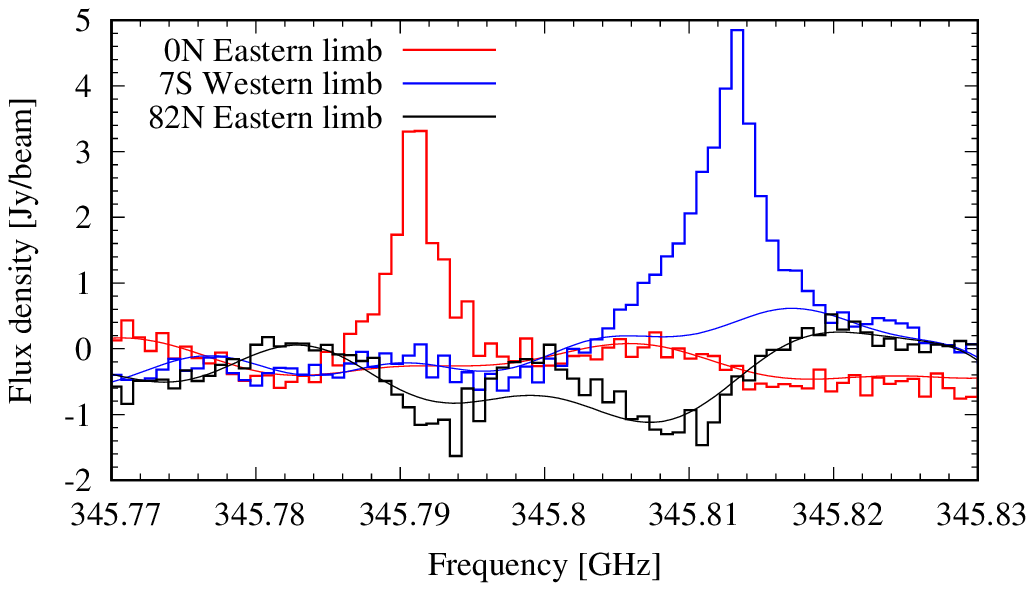}
    \caption{Example of three raw spectra obtained with SMA at the limb of Saturn and at various latitudes. The standing wave removal stage comprises the subtraction of sine waves and/or polynomials from the raw spectra. An example of standing wave fit is provided for each raw spectrum, with the corresponding color. }
    \label{fig:baselines}
  \end{figure}

  Passband calibration proceeded as follows. Initial correction of the bandpass at the full spectral resolution was done using the lower sideband spectral data (in a spectral line free region of both Saturn and Mars) which shared much of the spectral response of the analog filters with the upper sideband data, and applied to all sources in each sideband. Following that, 3C273 was used to further correct the data in the upper sideband, at significantly lower spectral resolution to partially compensate for the lower S/N per channel of that data. Unfortunately, the changing nature of the standing waves was not fully corrected with this method, and further processing was needed using the Saturn spectral data itself. A 4th-order polynomial, to remove broad continuum spectral slopes, as well as sine and cosine pairs of specific periodicity related to the standing waves, were fit to all the spectral data excluding the central window (where the CO line is located). The effect of this fitting is to remove roughly a linear slope across the CO window, along with periodic standing waves that are clearly seen in the other windows. While substantially improving the spectral fidelity of the central window, residual effects are still evident (see \fig{fig:baselines}).
  
\begin{figure}[!ht]
    \centering
    \includegraphics[width=0.7\textwidth]{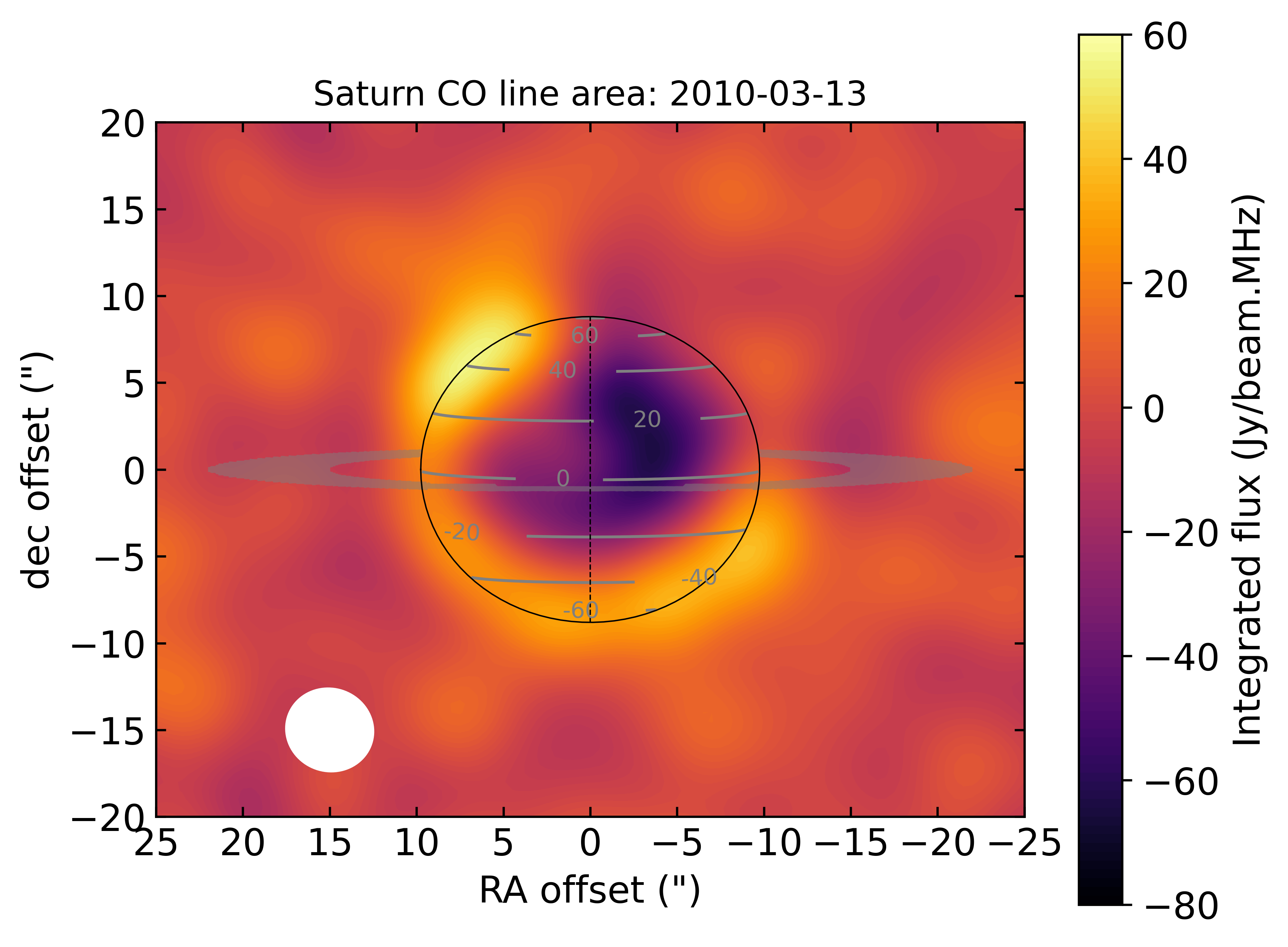}
    \caption{CO (J=3-2) line area map, as observed with SMA on March 13, 2010. The 1-bar level is shown with the black ellipse, the planet rotation axis is displayed with a dashed black line, and isolatitudes are indicated by gray contours. The expected position of the A and B rings is depicted by the gray filled area and the beam is illustrated with a white filled ellipse.}
    \label{fig:linearea-SMA}
  \end{figure}

  Spectral line imaging of faint lines over strong continuum sources is better achieved after fitting and removing the continuum in the visibilities. After continuum subtraction, the visibility data for the CO line spectral window were exported into AIPS (Astronomical Image Processing System) for imaging. The synthesized beam realized from the visibility data was 5.0'' $\times$ 4.7'' with a position angle of 71.4$^\circ$. Each spectral channel (812.5\,kHz width) was processed, providing an image of the continuum-subtracted CO emission in that channel, resulting in a spectral cube. The CO line is consistently detected in emission only at the planetary limb, where atmospheric path lengths are maximized, as demonstrated by the line integrated map in \fig{fig:linearea-SMA}. For spectral analysis of the CO line emission, we therefore only extract spectra along the limb, defined here as the 1-bar level. Given the spatial resolution of the observations, choosing any other level up to the upper limit of the stratosphere would imply no measurable change on the spectral lines, because the thickness of the atmosphere is negligible compared to the beam size. We choose to oversample the synthesized beam by a factor of 4, yielding 50 spectra along the Saturn limb to analyze. This enables to sufficiently oversample the latitudinal resolution of the observation. The SMA primary beam full width at half maximum (FWHM) of $\sim$35'' at the CO (J=3-2) frequency means limb emission is downweighted by a factor of 0.8. We consequently apply a primary beam correction of 1.25 to the flux densities of the continuum-subtracted spectra extracted at the limb from the spectral cube.
  
  \fig{fig:baselines} shows that the spectral quality is still hampered by residual standing waves incompletely removed during the passband calibration process. In some positions the standing waves have an amplitude on the order of that of the CO line (for example, at high northern latitudes). We remove those either using sine waves, or alternatively polynomial fits because the narrow bandwidth sometimes make it difficult to constrain sine wave fit parameters. The uncertainty on the CO line amplitude resulting from this data processing stage are $\sim$10\% at equatorial latitudes, 10-to-40\% at mid-to-high latitudes, even peaking at 100\% at high northern latitudes. These uncertainties are accounted for in the column density derivations hereafter.

  \subsection{January 2012 ALMA observations}
  We observed the CO (J=2-1) line at 230.538 GHz emanating from Saturn as part of project 2011.0.00808.S on January 9, 14, and 22, 2012, with 17, 16, and 18 antennas, respectively, in the Cycle 0 compact configuration. The baselines extended from 18\,m to 268\,m. Atmospheric precipitable water vapor amounted to 7\,mm, 4\,mm, and 0.5\,mm, respectively. The planet had an equatorial angular diameter of $\sim$17.1'' and a sub-Earth latitude of 15.1$^\circ$N. The observations were scheduled when Saturn's beacon was located at the planetary eastern limb\footnote{East/west direction indicated in this paper must be understood as planetary east/west, as opposed to sky east/west.} to increase S/N and limit apparent longitudinal smearing in the final images. Each observation consisted of a single pointing alternating between Saturn and the blazar 3C279 for calibration. The total on-source time was $\sim$55 minutes. Because the CO line is faint and lies on a strong continuum, additional 30 minutes of bandpass calibration observations were executed, but only on January 14 and 22. The lack of additional bandpass calibrator observations on January 9 significantly reduces the final line sensitivity in these observations, and we choose to discard them from the final coadd. All spectra were recorded with a spectral resolution of 244\,kHz.

  \begin{figure}[!ht]
    \centering
    \includegraphics[width=0.55\textwidth]{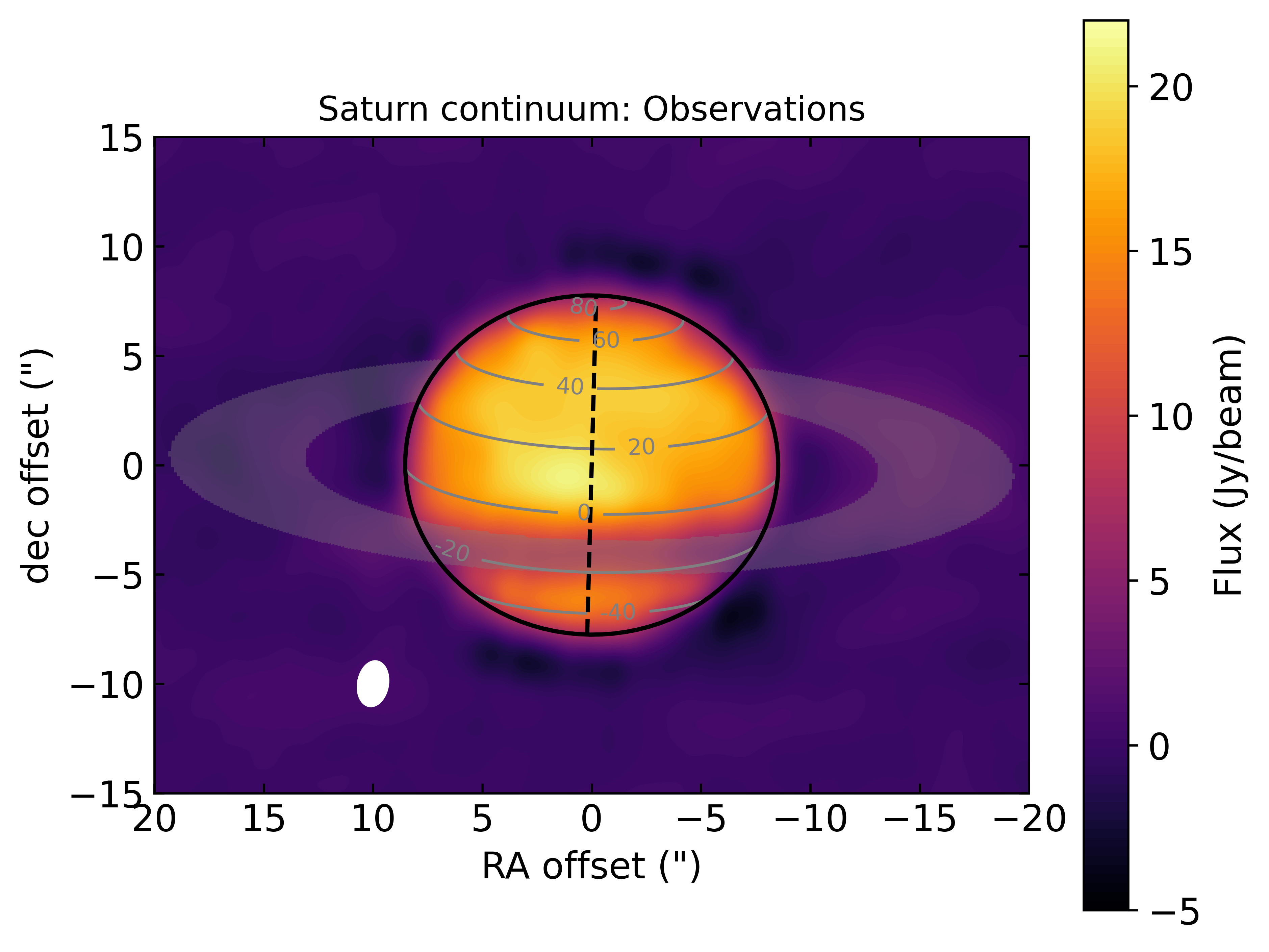}
    \includegraphics[width=0.55\textwidth]{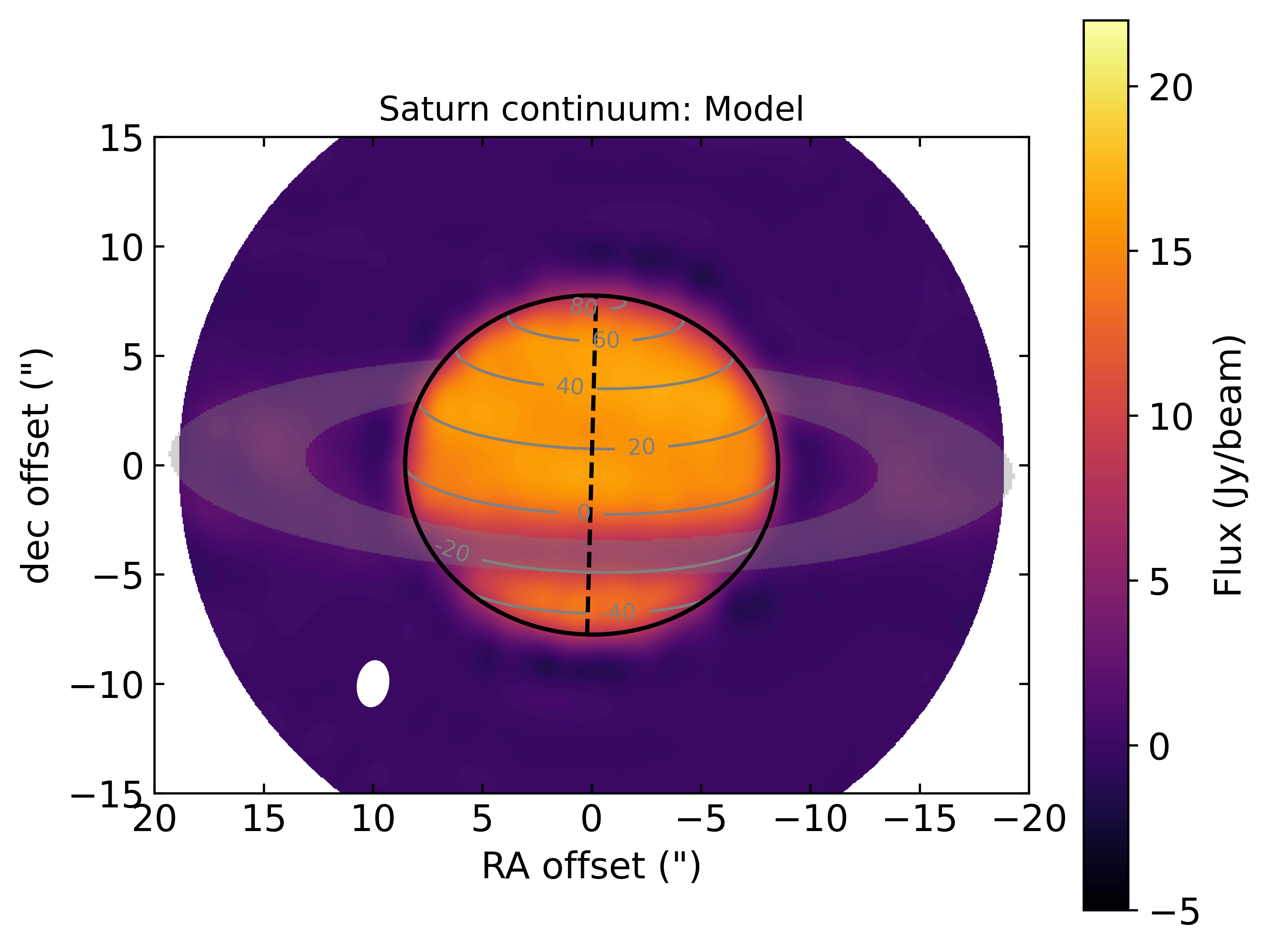}
    \includegraphics[width=0.55\textwidth]{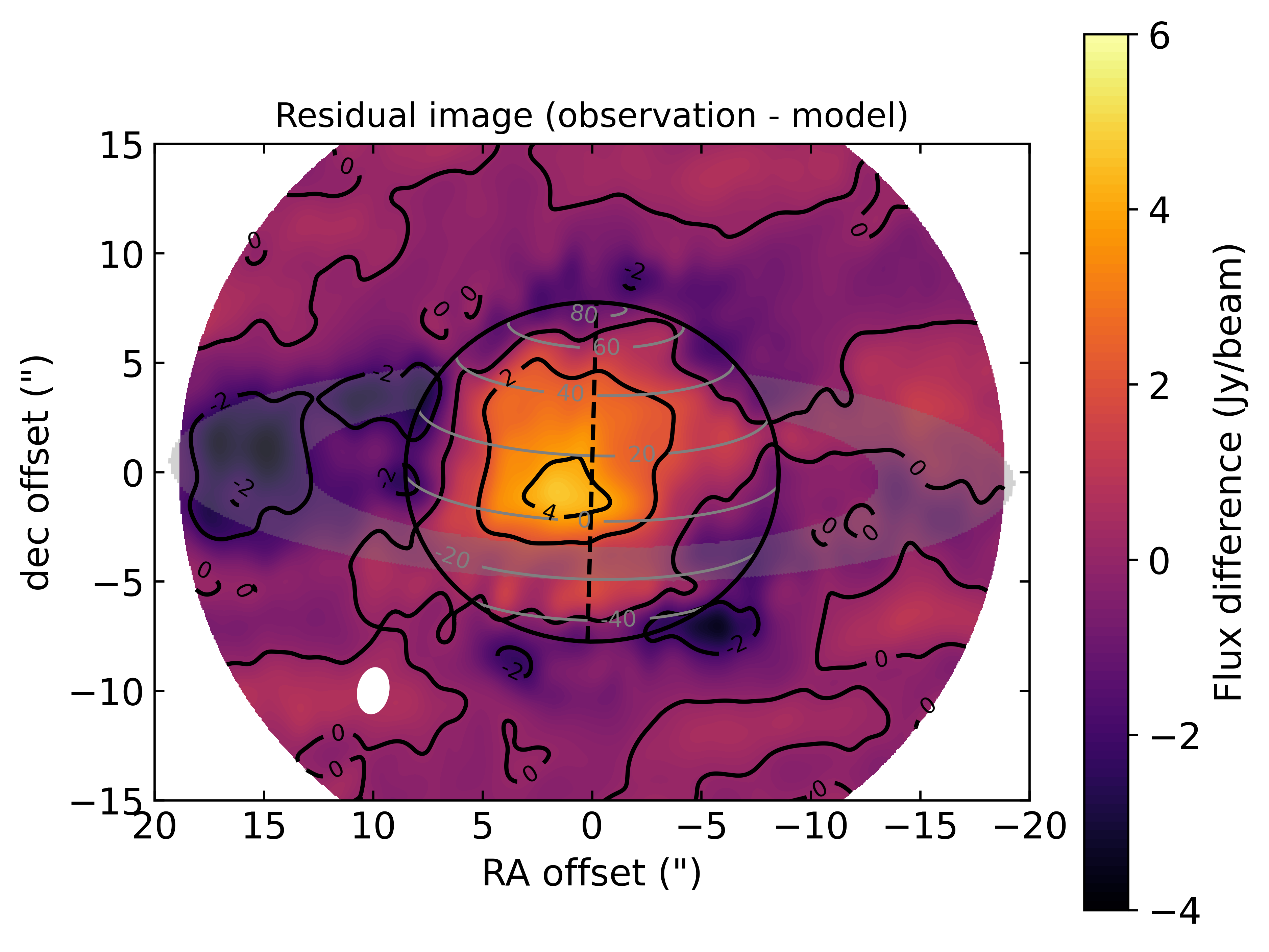}
    \caption{Saturn continuum flux density images at 230\,GHz. (Top) Saturn continuum image as observed with ALMA on January 14, 2012. (Center) Simulated continuum image in the conditions of the January 2012 observations. (Bottom) Residuals (observation - model) showing the overall good agreement between model and observations, especially at the limb. In all panels, the 1-bar level is shown with the black ellipse, the planet rotation axis is displayed with a dashed black line, and isolatitudes are indicated by gray contours. The expected position of the A and B rings is depicted by the gray filled area and the beam is illustrated with a white filled ellipse.}
    \label{fig:ALMA-cont}
  \end{figure}
  
  We reduced the data with the Common Astronomy Software Applications (CASA) version 4.7.2. After the calibration of the bandpass, gain, phase, and flux, we corrected the variation of Saturn's apparent size between January 14 and 22 by applying the relevant scaling factor on the uv distances in the January 22 data. There is a small difference remaining on total planet flux between the two dates: 978$\pm$14\,Jy on January 14 versus 951$\pm$12\,Jy on January 22, both after selfcalibration. We thus rescaled the January 22 amplitudes to obtain the same total planet flux as on January 14. This enabled us to concatenate the uv-tables from both dates into a single measurement set. We further processed the data to produce selfcalibrated continuum and continuum-subtracted spectral images. We used the multiscale algorithm in the clean process, with scales relevant to the features expected in the Saturn image (for example, 1'' for the limb spectral data, and larger scales for the continuum data). We cleaned the data within a combination of planet ellipse and ring ellipse masks for the continuum image, and only within a planet ellipse mask for the spectral cube, because there is no line emission in the rings. We chose the Clark algorithm to compute the synthesized beam. Similarly to the SMA data also, the continuum was subtracted from the visibilities to produce the continuum-subtracted spectral cube. The final spectral noise is 6.5\,mJy/beam per spectral channel.

\begin{figure}[]
    \centering
    \includegraphics[width=0.7\textwidth]{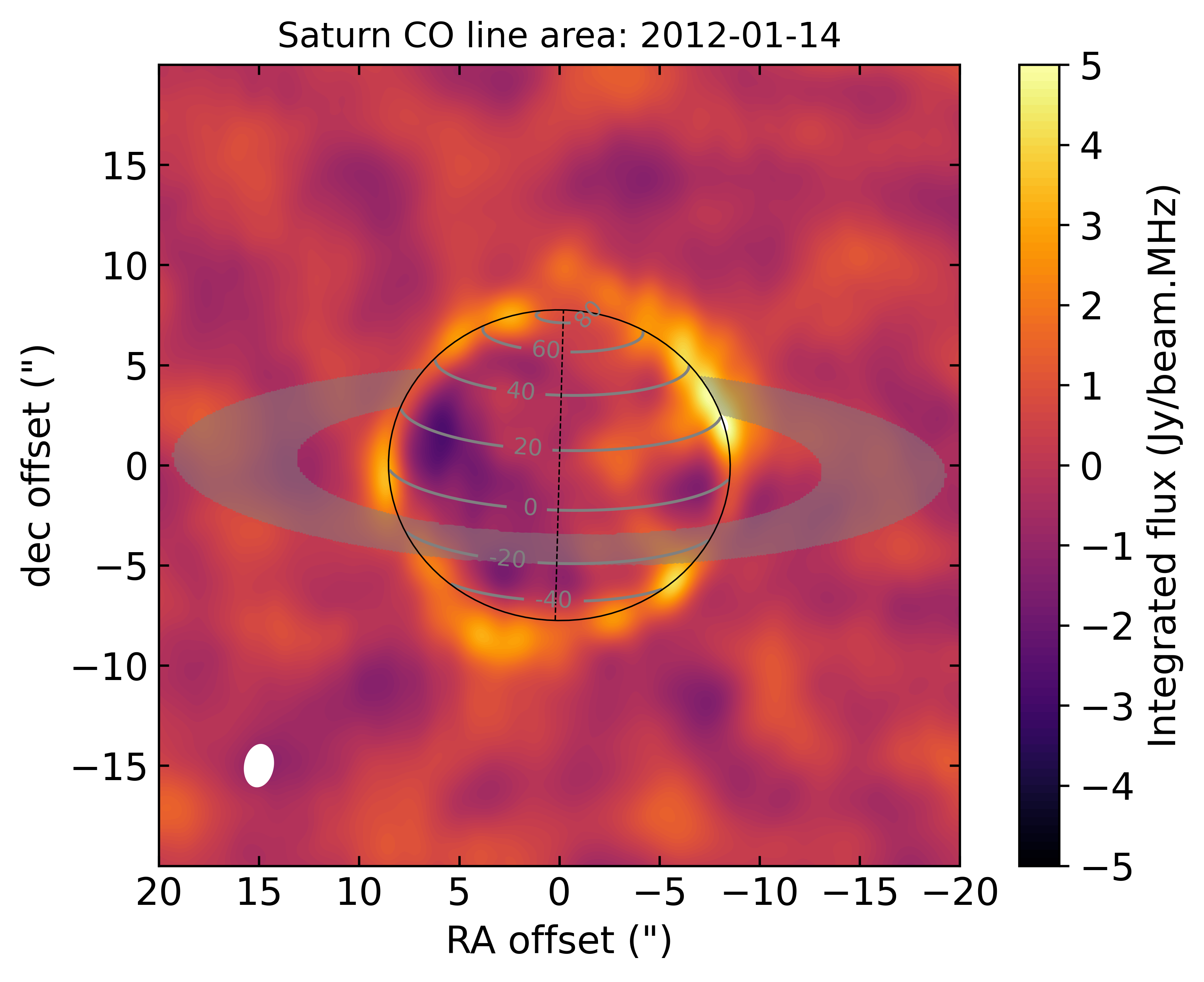}
    \caption{CO (J=2-1) line area map from the combined ALMA observations of January 14 and 22, 2012.  The 1-bar level is shown with the black ellipse, the planet rotation axis is displayed with a dashed black line, and isolatitudes are indicated by gray contours. The expected position of the A and B rings is depicted by the gray filled area and the beam is illustrated with a white filled ellipse.}
    \label{fig:linearea-ALMA}
  \end{figure}

  The compact configuration used for these observations resulted in a synthetic beam of 2.04'' $\times$ 1.34'' with a position angle of -9.9$^\circ$. Using the continuum image (see \fig{fig:ALMA-cont}), we find a residual pointing offset of +0.27'' in right ascension and +0.33'' in declination (i.e., 15-20\% of a synthetic beam) with respect to the center of the image. The resulting line area map shows that the CO line emission is concentrated at the limb (see \fig{fig:linearea-ALMA}), similarly to the SMA observations. We apply the relevant pointing correction before extracting the limb spectra, still applying a beam oversampling factor of 4, to improve the accuracy of the beam-convolved planet rotation subtraction in the wind measurement process (see Section \ref{sec:Winds}). We note that, contrary to the standing-wave-limited SMA data, the ALMA cube shows a $\sim$0.1\,Jy/beam CO absorption feature around equatorial latitudes within the planet disk and we discuss these in Section \ref{sec:ALMA-CO}.
  
\section{Models} \label{sec:Models}
  \subsection{Spectral modeling}
     \subsubsection{Thermal fields}
\begin{figure}[]
       \centering
       \includegraphics[width=0.7\textwidth]{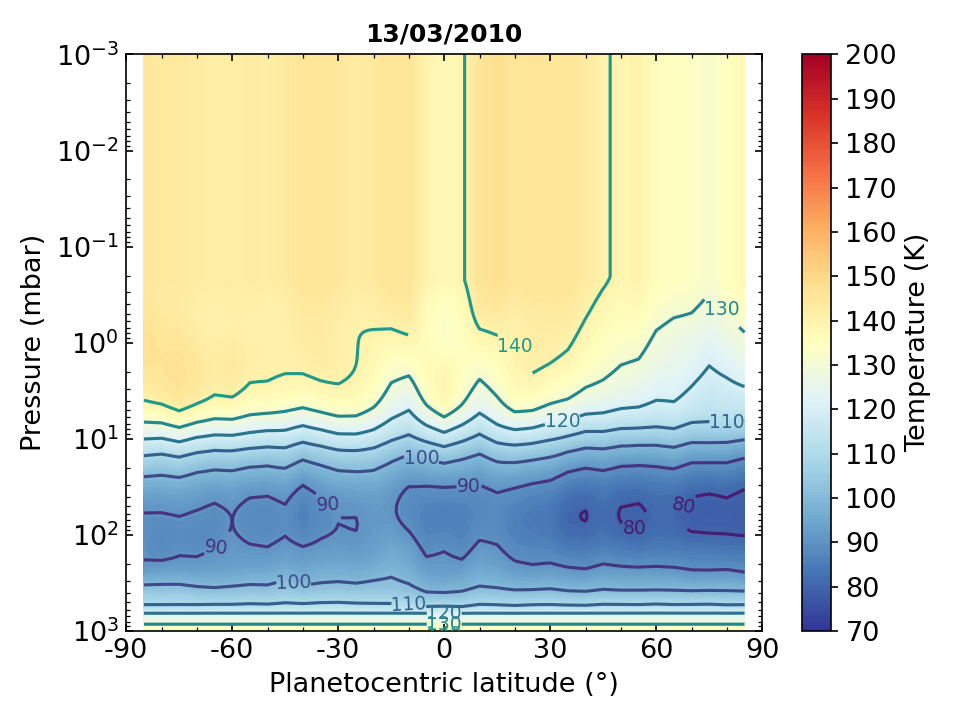}
       \caption{Zonal field of temperatures used for the analysis of the SMA data taken on March 13, 2010. The data are extracted up to 0.5\,mbar from the \citet{Fletcher2018b} dataset and expanded isothermally upward.}
       \label{fig:temps2010}
     \end{figure}

     For the SMA data from March 2010, we use the seasonal altitude-latitude thermal field retrieved from Cassini/CIRS measurements presented in \citet{Fletcher2018b}. The altitude-latitude field extracted for the exact same date is shown in \fig{fig:temps2010}. Similarly to \citet{Lefour2025}, we expanded the temperatures isothermally upward of the 0.2\,mbar level because of the lack of sensitivity of the nadir measurements used in the temperature retrievals. Consequently, we do not account for the thermal oscillation observed around equatorial latitudes \citep{Fouchet2008,Guerlet2018}. We discuss this point in Section \ref{sec:ALMA-CO}.
     
\begin{figure}[]
       \centering
       \includegraphics[width=0.7\textwidth]{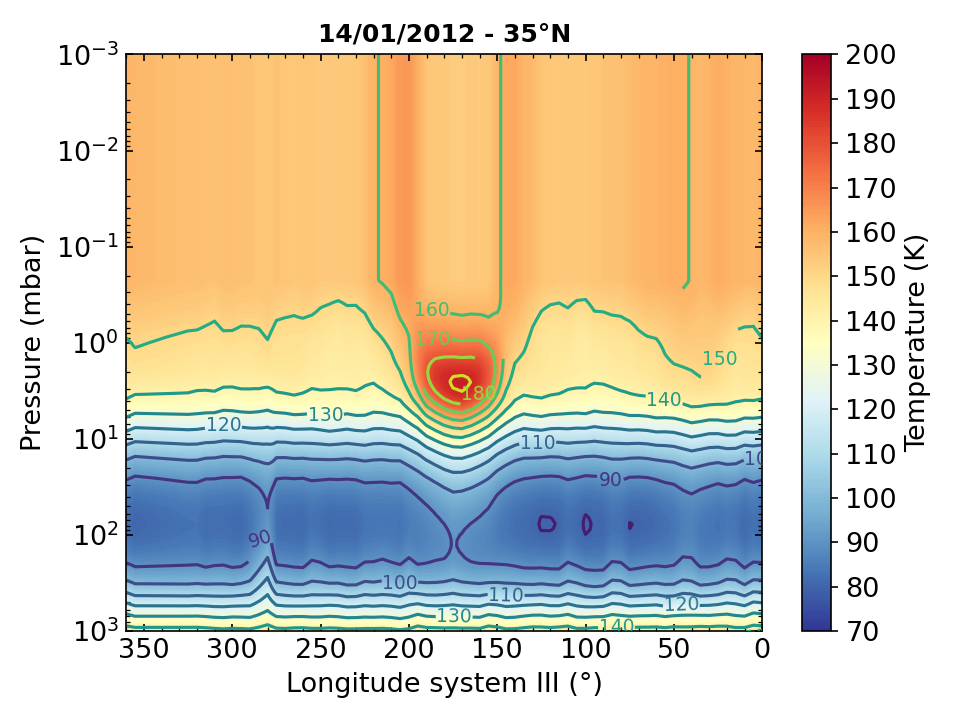}
       \includegraphics[width=0.7\textwidth]{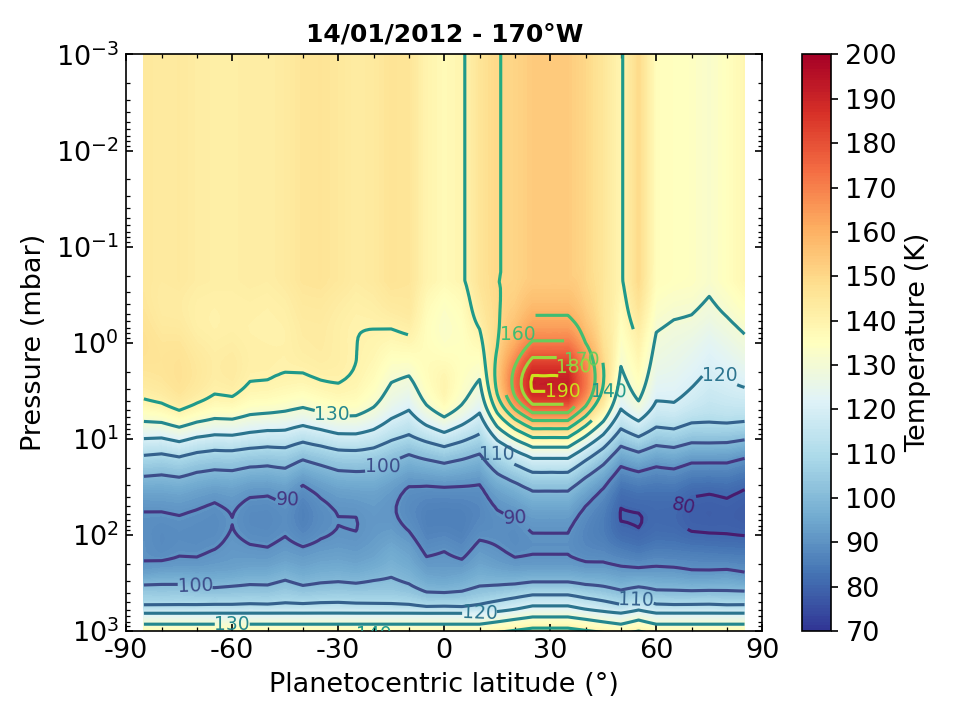}
       \caption{Temperature field at the beacon latitude (top) and longitude (bottom) in January 2012.}
       \label{fig:temps2012}
     \end{figure}

     For the ALMA data from January 2012, we combine two sources of data: (i) temperatures from \citet{Fletcher2018b} taken for all latitudes unaffected by the beacon, and (ii) temperatures retrieved as a function of altitude, latitude, and longitude from beacon-specific observations carried out on January 27, 2012, by \citet{Fletcher2012b}. We correct the beacon longitude to account for the beacon western drift \citep{Fletcher2012b} between January 14 and January 27 following \citet{Lefour2025}. Latitudinal and longitudinal cross sections of the final thermal field and taken at the location of the beacon are shown in \fig{fig:temps2012}. We use those data to analyze the spectra from January 2012. 

     \subsubsection{Background composition and CO distribution model}
     The background composition we use in the radiative transfer calculations is similar to those used in \citet{Cavalie2019} and \citet{Lefour2025} for what concerns H$_2$, He, CH$_4$, NH$_3$, and PH$_3$. These species all contribute to the continuum level of our observations. 
     
     For the CO vertical profile, we neglect any tropospheric CO \citep{Fouchet2017,Cavalie2024} in our calculations, because the tropospheric abundance of 1.2\,ppb only produces a negligible and very broad absorption that we cannot capture with the limited bandwidth and bandpass quality of the data. Also, because of the limited bandpass quality of the SMA data, and to a less pronounced level, of the ALMA data, we cannot fully rely on the broad wing shape of the lines after the baseline ripple removal stage. We thus cannot reliably constrain the CO vertical profile. As a consequence, we choose to adopt the physical profile of \citet{Cavalie2010} derived from atmospheric transport following a comet impact occurring $\sim$220\,years before the observations, and to scale it to fit the spectra at each limb position so as to estimate the CO abundance at the pressure where the contribution functions peak for the two lines (see \fig{fig:contribution}).

     \subsubsection{Radiative transfer model}
\begin{figure}[!t]
     \centering
     \includegraphics[width=0.7\textwidth]{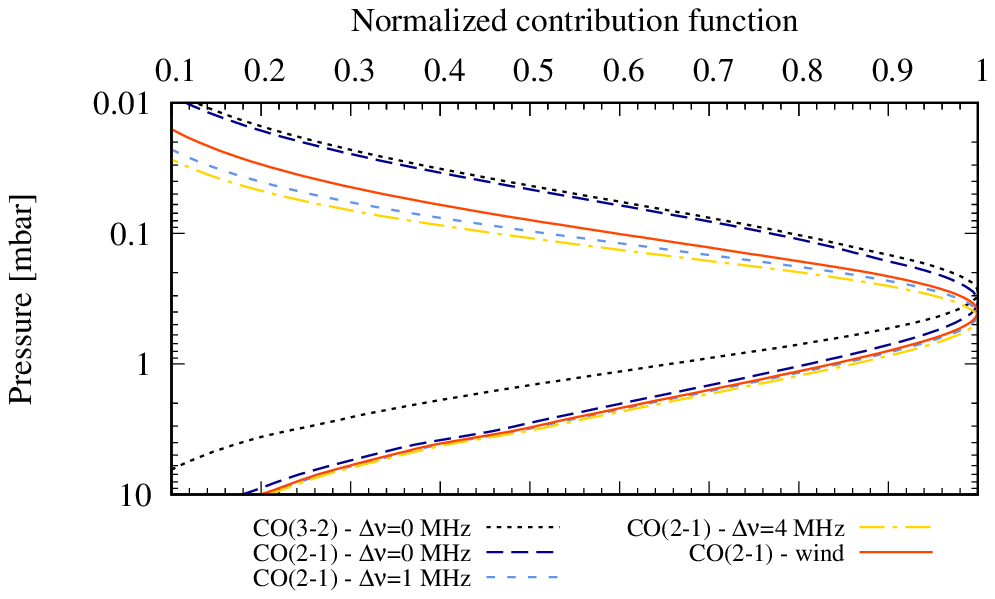}
     \caption{Normalized contribution functions at the line center for the CO (J=3-2) line observed with SMA in March 2010 (dotted black line) and CO (J=2-1) line observed with ALMA in January 2012 (longdashed, dark blue line). Contribution functions are also presented for the CO (J=2-1) line for 1\,MHz (shortdashed, light blue line) and 4\,MHz (dashed-dotted, yellow line) frequency offsets with respect to the line center. The wind contribution function pertaining to the CO (J=2-1) observations is plotted with a solid red line. All contribution functions are calculated at the equatorial planet limb with the rescaled 220-year-old-comet-impact profile of \citet{Cavalie2010} and account for the relevant spatial and spectral convolutions. }
     \label{fig:contribution}
   \end{figure}

     We use the line-by-line radiative transfer model of \citet{Cavalie2019}, which accounts for the 3D ellipsoidal shape of Saturn. The continuum is mainly caused by the collision-induced absorption spectra of H$_2$-H$_2$, H$_2$-He, and H$_2$-CH$_4$ pairs. In this respect, we adopt the routines prescribed by \citet{Borysow1985,Borysow1988} and \citet{Borysow1986}. The continuum is also affected by the broad wings of NH$_3$ and PH$_3$ lines \citep{Moreno1998}. The spectroscopic constants of the NH$_3$, PH$_3$, and CO lines modeled in this paper have been extracted from the JPL Spectroscopy database. We have computed the line broadening parameters following \citet{Fletcher2007} for NH$_3$, \citet{Levy1993,Levy1994} for PH$_3$, and \citet{Dick2009} for CO. The adopted values are summarized in Table \ref{tab:spectro}. With the adopted CO vertical profile, we find that the CO (J=2-1) and (J=3-2) lines at the limb have an opacity of 0.36 and 1.15 at the line center and at the equator, and that they are produced in the 0.1-1 \,mbar pressure range, with the line center contribution function peaking at $\sim$0.3\,mbar (see \fig{fig:contribution}).

     All spectral analyzes in this paper are done on the continuum-subtracted data cubes. We have thus checked that our continuum model, especially in the vicinity of the limb is consistent with the observations within calibration uncertainties (see \fig{fig:ALMA-cont}). The main flux density differences are found within the disk and have no consequence on the results presented hereafter. There are two regions at the limb where differences up to 2\,Jy/beam are found. They are both located where the rings influence the continuum budget, one on the western limb between 20 and 40\degre N and the other on the eastern limb between 40 and 50\degre S. The first one results from imperfect cleaning of the rings. They appear less obvious in the image on the western side compared to the eastern one. The observation-model residual image also shows a better match between model and observation for the rings on the eastern side. However, the planet line and continuum remain consistent between the two limbs as we consistently find compatible CO abundances on both limbs as is shown in Section \ref{sec:ALMA-CO}. The other region, between 40 and 50\degre S on the eastern limb, must result from limitations in the data and/or cleaning. It translates to difficulties in fitting the data, as discussed further in Section \ref{sec:ALMA-CO}.

     \begin{table*}[!t]
         \caption{Line broadening parameters adopted for the NH$_3$, PH$_3$, and CO lines considered in this paper.}
         \label{tab:spectro}
         \centering
         \begin{tabular}{lllll}
            \hline
            Lines & $\gamma$ (cm$^{-1}$.atm$^{-1}$) & $n$ & $T_\text{ref}$ (K) & Reference \\
            \hline
            NH$_3$ (all) & 0.072 & 0.73 & 296 & \citet{Fletcher2007} \\
            
            \hline
            PH$_3$ (J=1-0) & 0.1019 & 0.71 & 296 & \citet{Levy1993,Levy1994} \\
            PH$_3$ (J=2-1) & 0.1006 & 0.71 & 296 & \\
            \hline
            CO & 0.067 & 0.59 & 296 & \citet{Dick2009} \\
            \hline
         \end{tabular}
     \end{table*}
     
     The model also includes the emission and absorption of the main rings. All details relative to the various ring brightness temperatures and absorption coefficients can be found in \citet{Cavalie2019} (and references therein).
     
  \subsection{Wind retrieval model}
  We use the wind retrieval technique developed by \citet{Cavalie2021} for ALMA observations of Jupiter. It was also applied to the May 2018 ALMA observations of Saturn \citep{Benmahi2022}, and it uses a parametrized analytical function to fit the observed lines and derive the wind-induced spectral shift. We apply it to the ALMA data from January 2012 to assess the stratospheric wind regime during the storm and when the beacon was active. On the other hand, we choose not to apply it to the SMA data owing to the continuum ripple limitations. Their subtraction results in much too large uncertainties (several 100 m.s$^{-1}$) for a meaningful assessment of the stratospheric winds in 2010. 
   
\begin{figure}[!t]
     \centering
     \includegraphics[width=0.7\textwidth]{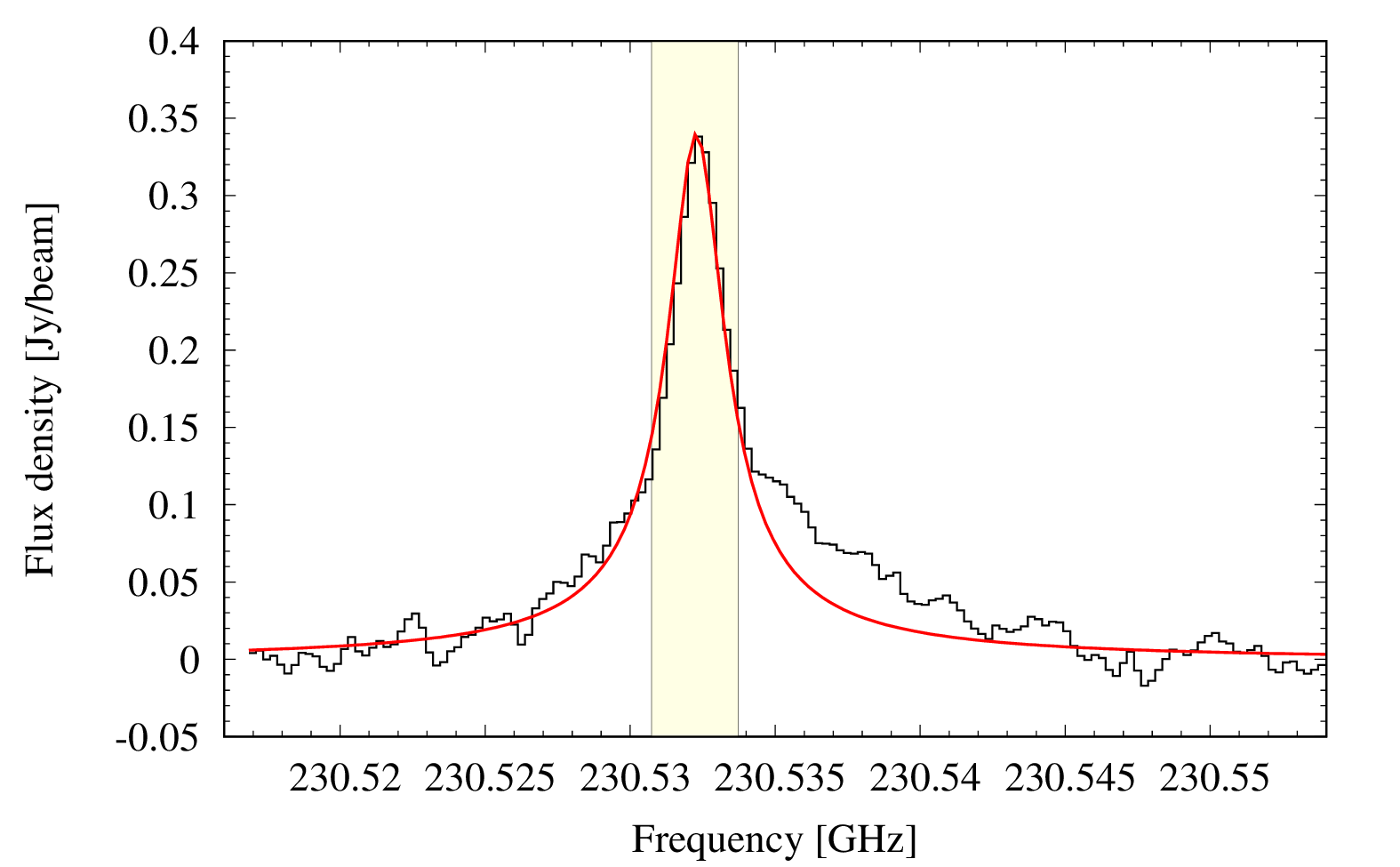}
     \caption{Spectral line observed with ALMA at the eastern limb and at 35$^\circ$N, in the beacon. Because the line is asymmetrical in its broad wings (see explanation in the text), the line is fitted with the wind retrieval algorithm (and its symmetrical analytical function parametrized following \citealt{Benmahi2022}) only within $\pm$1.5\,MHz of the line center, to avoid an erroneous wind speed retrieval. The line is shown in black and the fit in red, with the fitted interval highlighted with the light yellow box. }
     \label{fig:wind-fit}
   \end{figure}

  The main difference with the \citet{Benmahi2022} data results from the more limited spatial resolution of the 2012 ALMA data. With a $\sim$1.7'' beam, we have a planet spatial resolution (planet size/beam size; PSR) of about 10, when the 2018 observations of \citet{Benmahi2022} had a PSR of $\sim$40. This lower PSR results in mixing the contributions of disk and limb line-of-sights (LOS) that have different LOS-projected planet rotation velocities in a beam centered on the limb. The lines as observed at the limb are thus broader and, even more importantly, less symmetric than in the \citet{Benmahi2022} data. After careful assessment, we thus choose to apply the wind retrieval algorithm only to a frequency range comprised within $\pm$1.5\,MHz of the line peak. This enables to have sufficient data points for the fit while using only the more symmetric part of the observed line. An example is shown for the spectrum extracted at 35$^\circ$N on the eastern limb (see \fig{fig:wind-fit}).
  
\section{CO meridional distribution} \label{sec:Meridional}
   \subsection{Prestorm meridional distribution}
\begin{figure}[!t]
     \centering
     \includegraphics[width=0.7\textwidth]{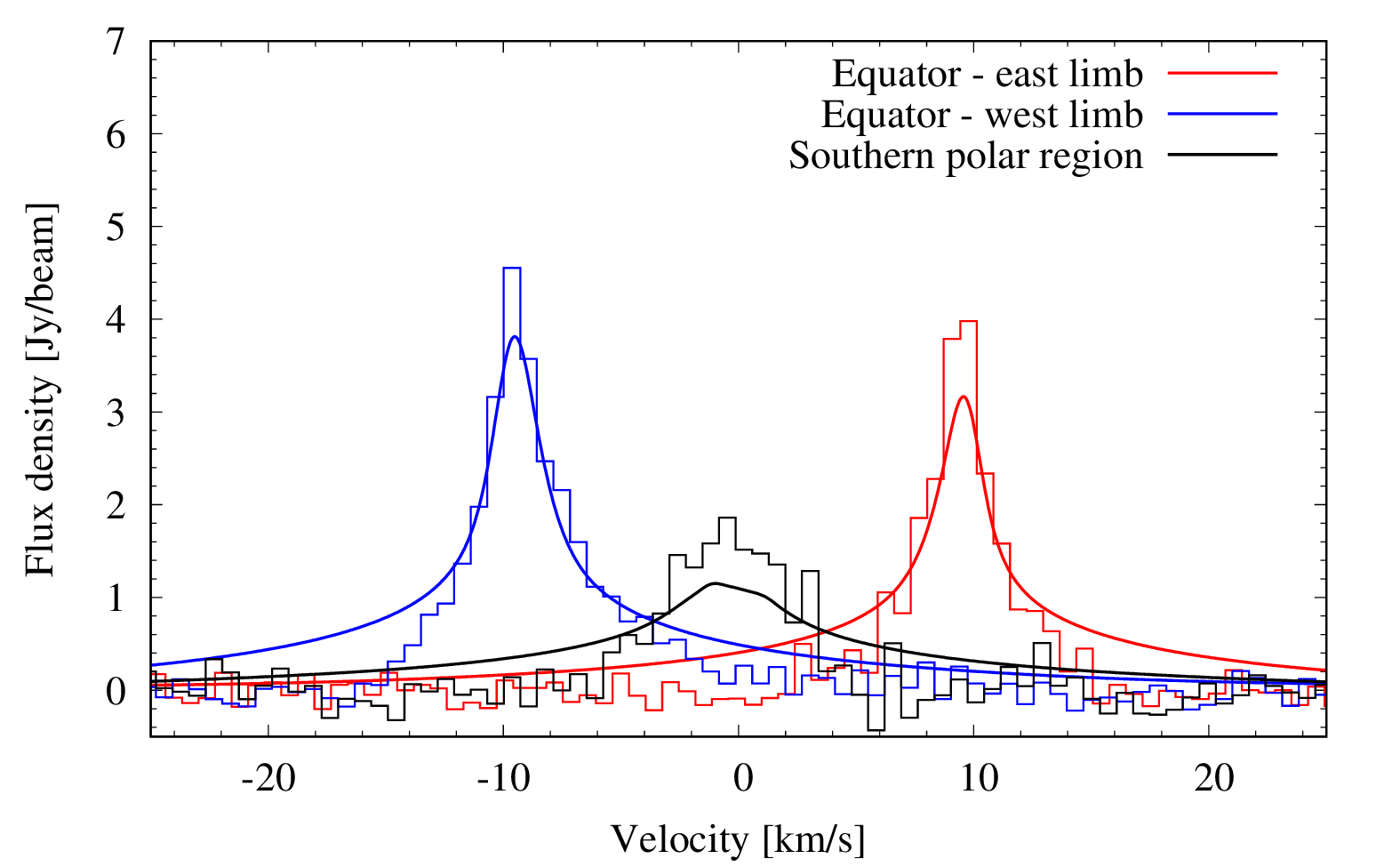}
     \caption{Example of a final spectra obtained with SMA in March 2010 and corresponding best fits obtained for a few latitudes, and exploring the two limbs with the 220-year-old-comet-impact profile of \citet{Cavalie2010} after proper rescaling. The lines are plotted in the velocity frame with respect to the line rest frequency (345.796\,GHz). They are redshifted on the eastern limb and blueshifted on the western limb because of the planet rapid rotation.}
     \label{fig:spectre_CO_SMA}
   \end{figure}

   The 2010 SMA data are generally fit quite poorly with the rescaled CO vertical profile, as shown with a few examples in \fig{fig:spectre_CO_SMA}. This is undoubtedly a result of the limitations induced by the significant processing required to remove baseline ripples, which contributed to remove the broad component of the line wings. We can then only derive a mean CO mole fraction at the 0.3\,mbar level, i.e., where the contribution function peaks, of (1.7$\pm$1.2)\dix{-7}, using the pointings for which the best fit for a given pointing on the limb satisfies $\chi^2/N$$<$9, where $N$ is the number of spectral data points used for the $\chi^2$ computation\footnote{For SMA data, $N$ is taken as 30, so that $\chi^2$ is computed over $\sim$25\,MHz. For ALMA data, $N$ is taken as 100, because the spectral resolution is about four times better.}.    
         
   \subsection{CO meridional distribution in January 2012 \label{sec:ALMA-CO}}
   There is no detectable CO feature in the ALMA spectral cube except at the limb in emission and on the disk around equatorial latitudes in absorption. While the limb data are analyzed in the next paragraphs, we briefly discuss the disk center spectra.
   
\begin{figure}[!t]
     \centering
     \includegraphics[width=0.7\textwidth]{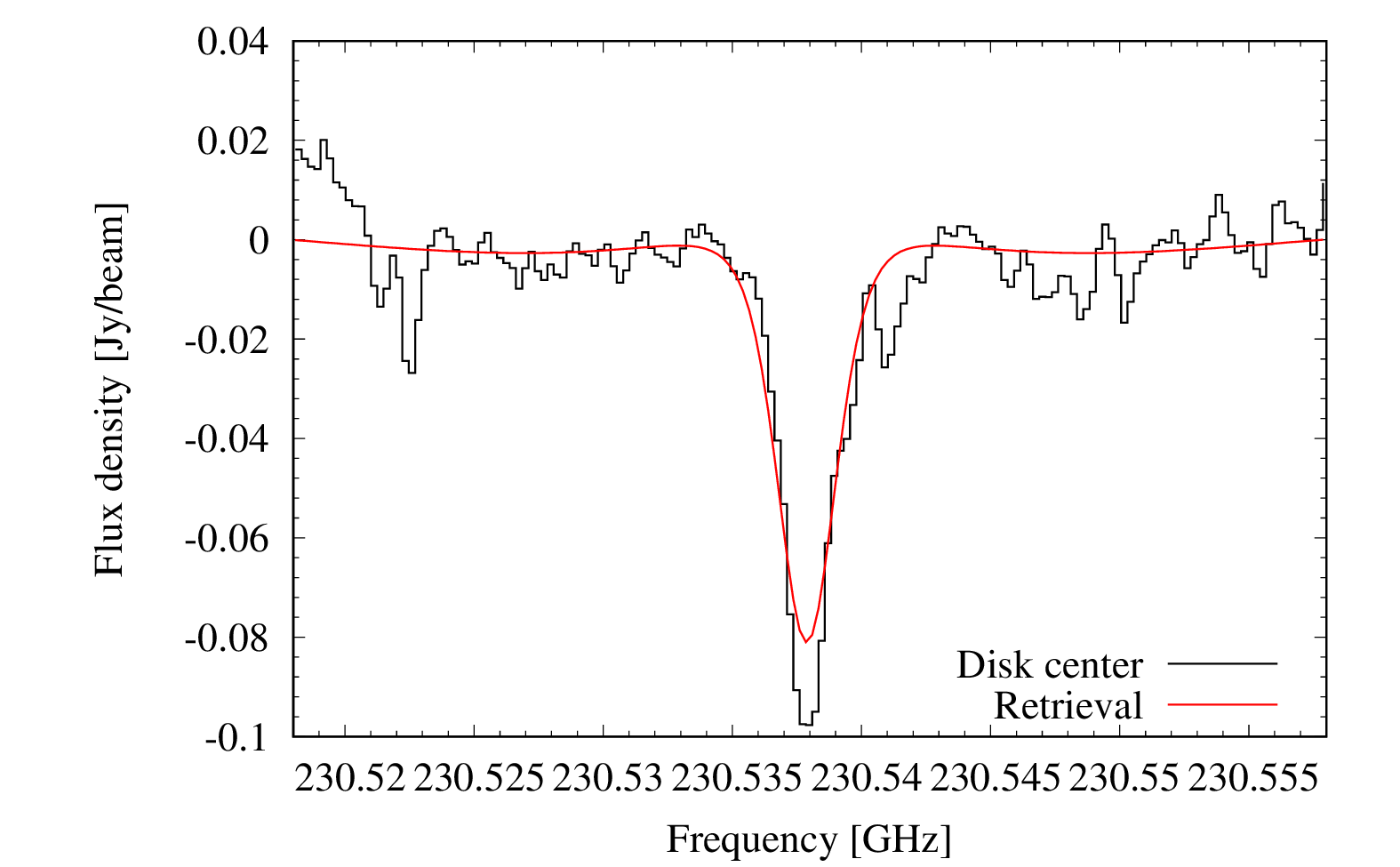}
     \includegraphics[width=0.7\textwidth]{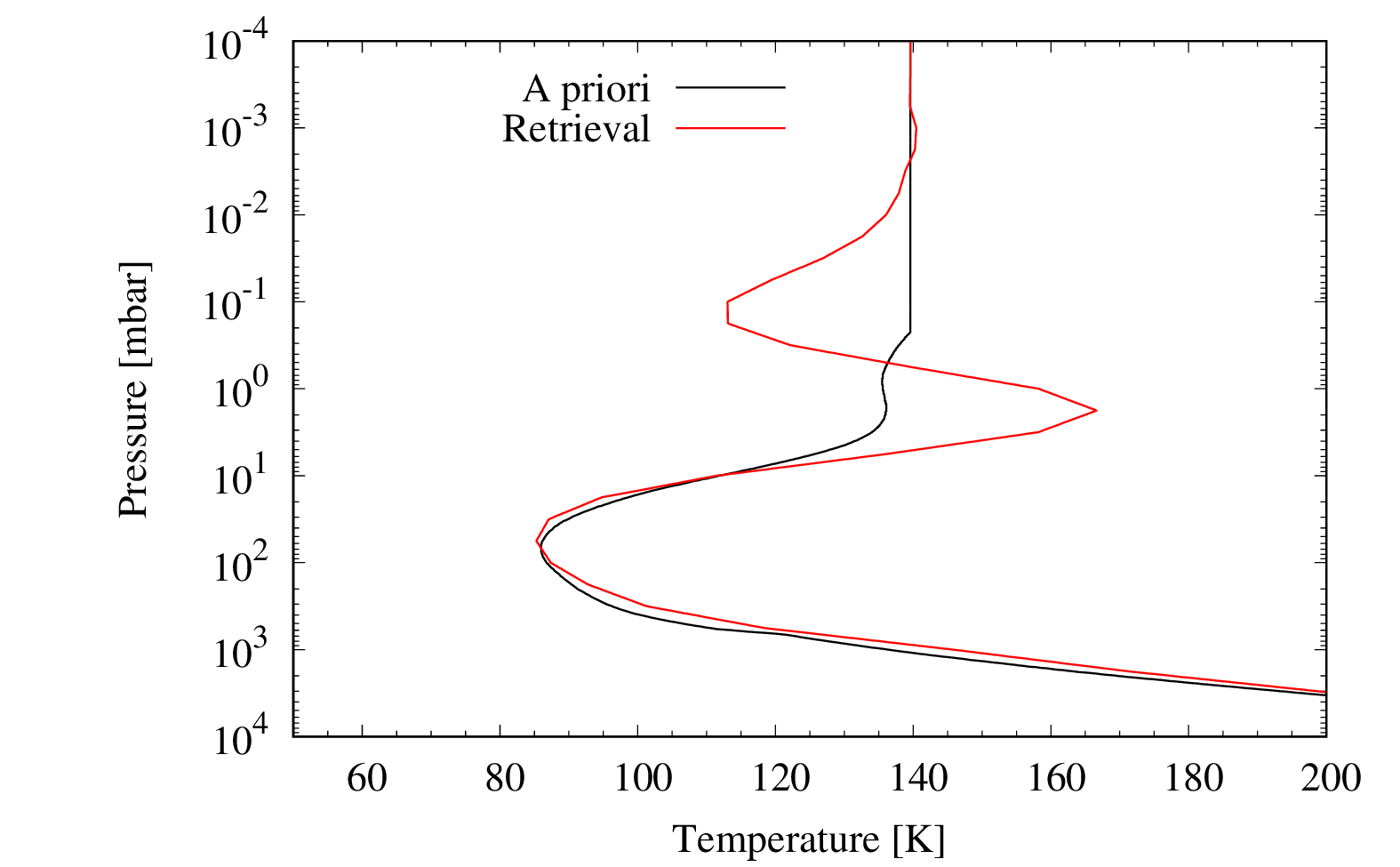}
     \caption{Disk-center spectrum and retrieved temperature profile demonstrating the strong thermal inversion centered around 0.3\,mbar required to produce the narrow absorption feature observed at equatorial latitudes.}
     \label{fig:SAO_retrieval}
   \end{figure}

   An absorption feature of $\sim$0.1\,Jy/beam is consistently detected at equatorial and near-equatorial latitudes from east to west. The FWHM of the line is $\sim$2\,MHz, which, given the collisional halfwidths, implies that the line is formed around the 1\,mbar level. Such absorption can consequently be caused only by a local thermal inversion, i.e., a locally negative temperature gradient in the stratosphere. The equatorial latitudes of Saturn are known to host a semiannual oscillation (SAO) of the stratospheric temperatures \citep{Fouchet2008}. A thermal inversion of $\sim$10\,K in the 0.1-1.0\,mbar pressure range was indeed derived from Cassini/CIRS observations by \citet{Guerlet2011} at the relevant latitudes in February 2010, i.e., about 2 years before the ALMA observations. We can qualitatively produce a CO absorption using such temperature vertical profile, but the depth of the model line is vastly underestimated. Using the retrieval algorithm of \citet{Fouchet2008} coupled with our radiative transfer model, we can better constrain the amplitude of the temperature oscillation and we find that it must be on the order of $\sim$50\,K between a maximum at 2\,mbar and a minimum at 0.1-0.2\,mbar (see \fig{fig:SAO_retrieval}). Even if a thermal oscillation on the order of 40\,K with an opposite phase was reported from 2015 data by \citet{Guerlet2018}, such a quantitative assessment from the ALMA disk-center data should be taken with caution. The lack of short spacings in the interferometric visibilities inherently filters out a significant fraction of the total flux in the disk-center data. Consequently, the shape of the spectrum remains quite uncertain. This may be why we fail at better fitting the line. A proper analysis of disk-center data would require dedicated observations with, for example, the Atacama Compact Array (ACA). In what follows, we thus focus our quantitative analysis of the ALMA observations to the limb data. 
   
\begin{figure}[!t]
     \centering
     \includegraphics[width=0.7\textwidth]{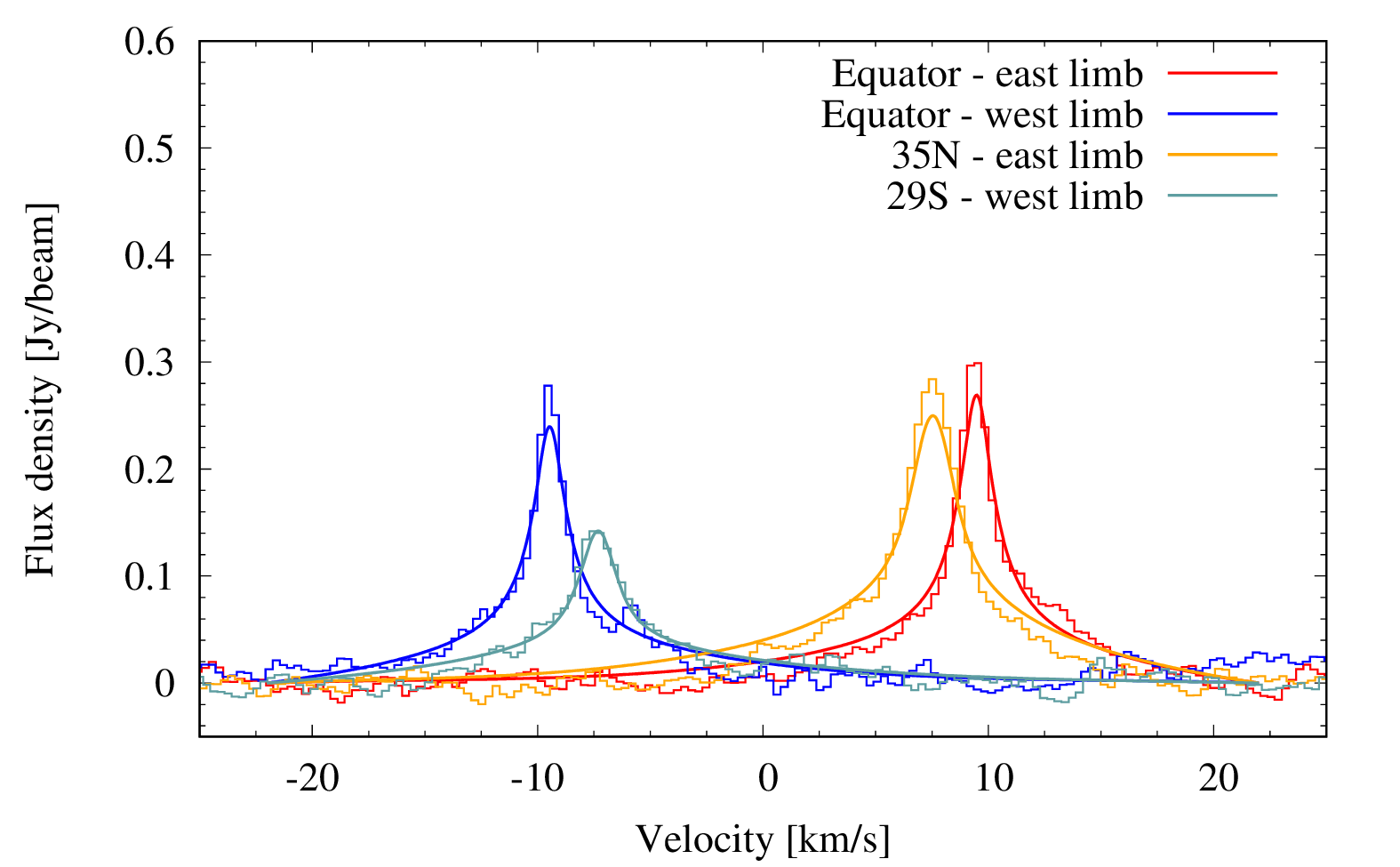}
     \caption{Example of final spectra obtained with ALMA in January 2012 with their respective best fits for various latitudes exploring the two limbs, with a highlight on the location of the beacon. The strongest emission seen in the beacon (for example, orange spectrum and fit) is caused by the temperature increase at mbar pressures inside the hot vortex. The lines are plotted in the velocity frame with respect to the line rest frequency (230.538\,GHz). }
     \label{fig:spectre_CO_ALMA}
   \end{figure}

\begin{figure}[!t]
     \centering
     \includegraphics[width=0.7\textwidth]{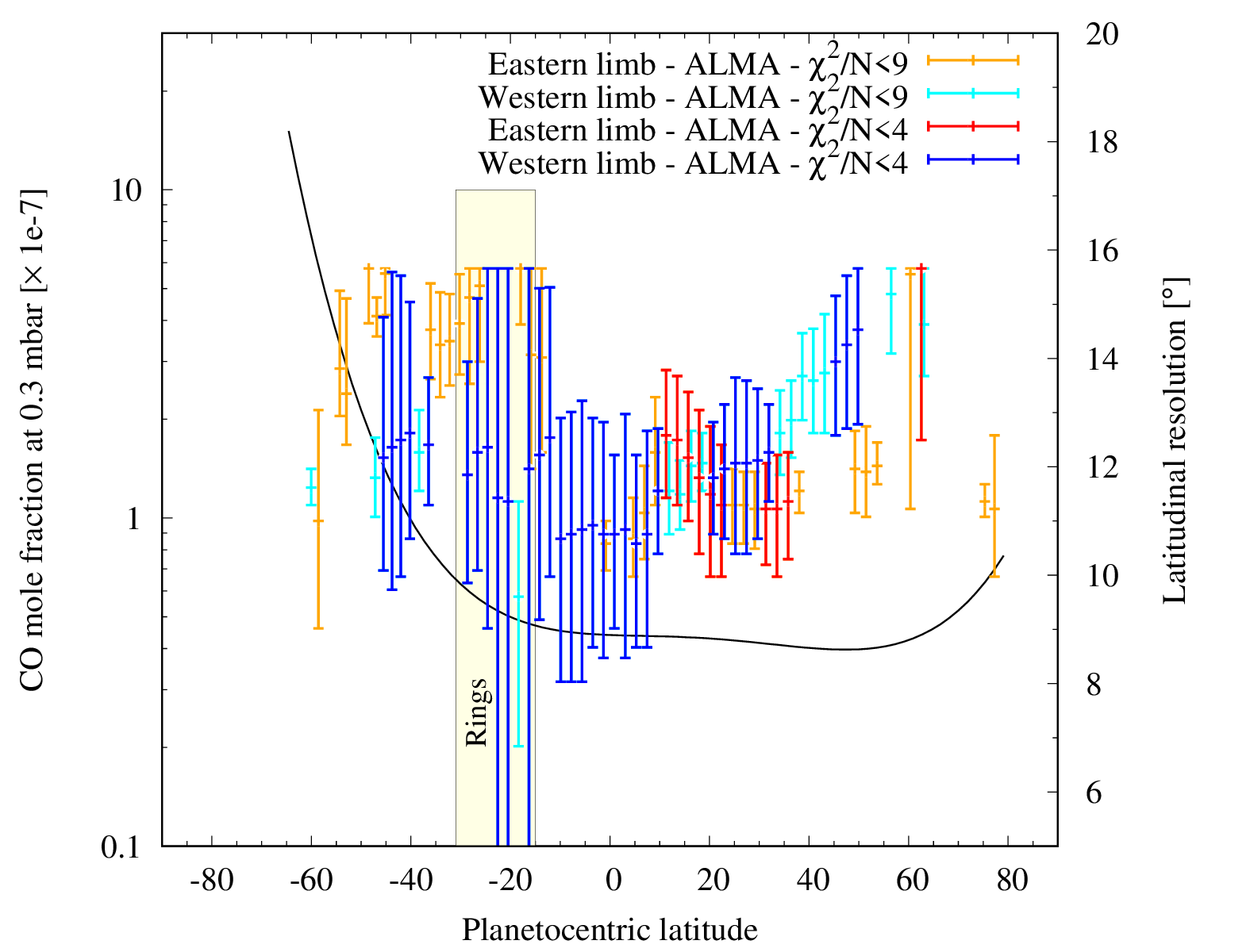}
     \caption{CO mole fraction measured at 0.3\,mbar in Saturn's stratosphere in January 2012 with ALMA as a function of latitude. The mole fractions are derived from fits made at the planet eastern (red line) and western (blue line) limbs with the 220-year-old-comet-impact profile of \citet{Cavalie2010} properly rescaled at each latitude to best fit the data. Error bars indicate the ranges of columns that fit the data with $\chi^2/N$$<$4 (red line for the eastern limb and blue line for the western limb) and $\chi^2/N$$<$9 (orange line for the eastern limb and cyan line for the western limb). The latitude region affected by ring shadowing in the 2012 ALMA data is highlighted with the light yellow box. The latitudinal resolution as a function of latitude is plotted in solid black on the second y-axis.}
     \label{fig:CO_ALMA}
   \end{figure}

  The 2012 ALMA data are better reproduced than the SMA data with the rescaled comet profile. As can be seen on the selected spectra of \fig{fig:spectre_CO_ALMA}, it is still noticeable that they also suffer, yet at a lower level than the SMA spectra, from the baseline removal process, because the broad wings are still not very well reproduced. The best fits are reflective of a compromise that produces too faint a line core and too much signal in the broad wings. The CO mole fraction at 0.3\,mbar is (1.7$\pm$0.7)\dix{-7}, when using the pointings for which the best fit satisfies $\chi^2/N$$<$9 and excluding those affected by ring shadowing. When using a more restrictive criterion of $\chi^2/N$$<$4 to compute the mean CO mole fraction at 0.3\,mbar, we find (1.4$\pm$0.8)\dix{-7}. This is only marginally compatible with \citet{Cavalie2010} who had 0.6\dix{-7} nominally at this pressure. The meridional distribution of the CO abundance at 0.3\,mbar is displayed in \fig{fig:CO_ALMA}, and it is compatible with a homogeneous distribution. The mean value derived from the SMA dataset is compatible with the ALMA meridional profile. 
    
   Finally, we find no significant change in the region of the storm, when comparing the lines on the eastern and western limbs at 20-50$^\circ$N. Even the line width is not significantly affected. This seems contradictory with what could have been expected from the downwelling winds that push material deeper, as reported by \citet{Moses2015} to explain the hydrocarbon abundances observed inside the beacon \citep{Fletcher2011,Fletcher2012b,Hesman2012} and \citet{Lefour2025} from H$_2$O observations.

\section{Stratospheric winds during the storm} \label{sec:Winds}
\begin{figure*}[!t]
     \centering
     \includegraphics[width=0.95\textwidth]{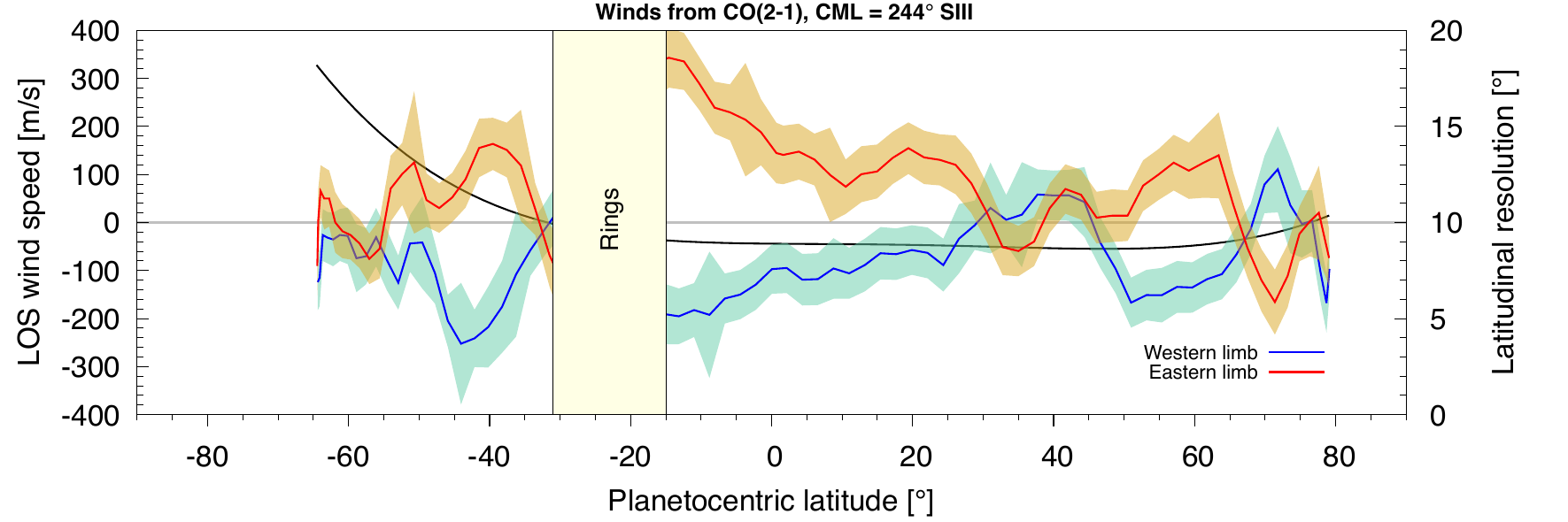}
     \caption{Line-of-sight Doppler winds in Saturn's stratosphere retrieved around 0.5-1\,mbar on January 14 and 22, 2012, from the ALMA CO (J=2-1) mapping observations. The eastern and western limb velocities (left axis) are plotted as a function of latitude in red and blue, respectively, with 1-$\sigma$ error bars. The region obscured by the rings and from which we do not retrieve wind speeds is masked by a light yellow box. The black line shows the variation of the latitudinal resolution as a function of latitude (right axis). CML stands for Central Meridian Longitude, and SIII stands for System III. Note the symmetry between the east and west limb wind speeds, demonstrating the robustness of our measurements up to 80$^\circ$N.}
     \label{fig:winds_summary}
   \end{figure*}

\begin{figure*}[!t]
     \centering
     \includegraphics[width=0.95\textwidth]{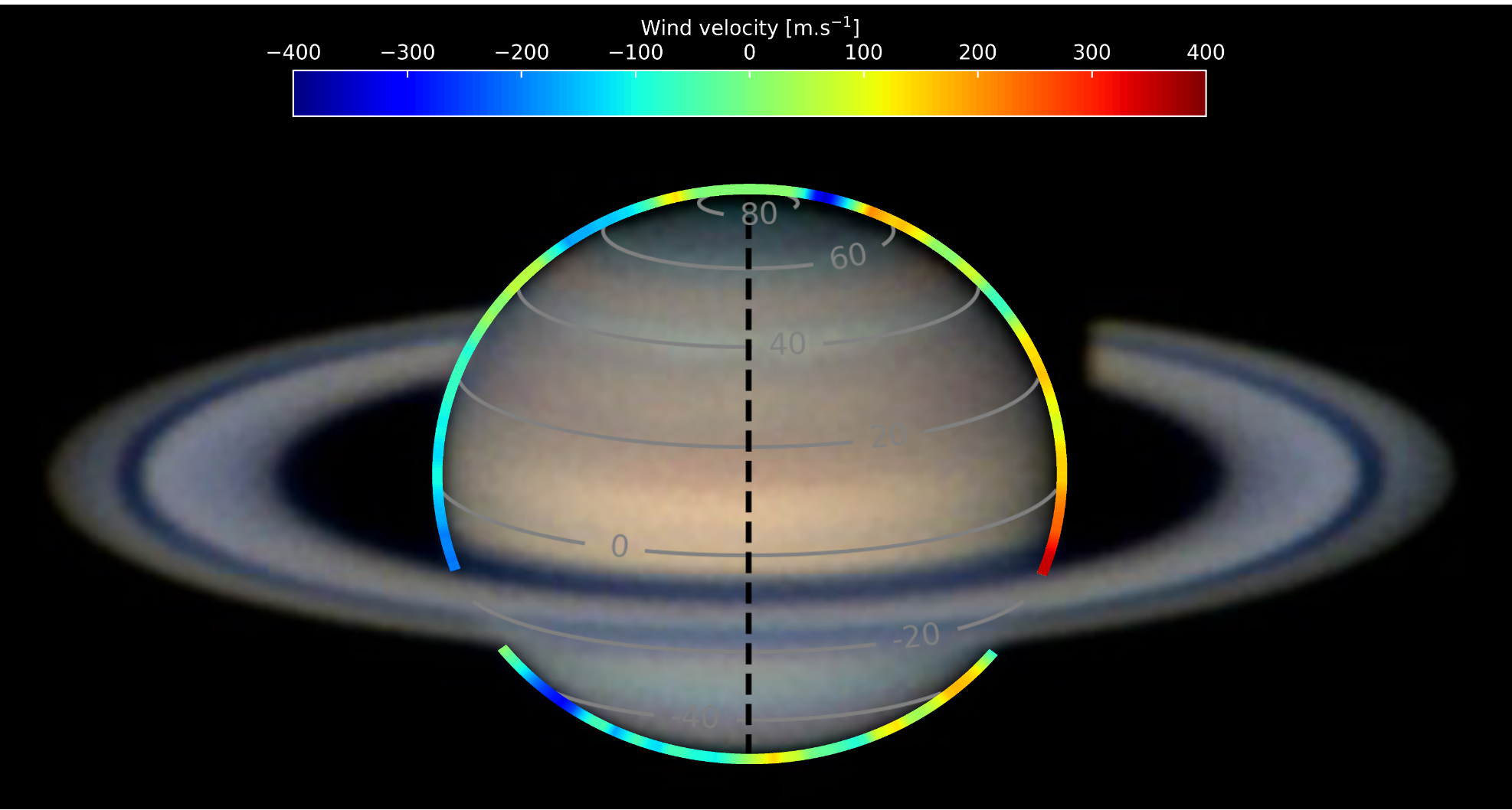}
     \caption{Zonal wind map in Saturn's stratosphere retrieved around 0.5-1\,mbar on January 14 and 22, 2012, from ALMA CO (J=2-1) observations. The wind speeds have been corrected for line-of-sight projection effects. The planet rotation axis is displayed with a dashed black line, and isolatitudes are indicated by gray contours. The background image of Saturn was captured by amateur astronomer F. Willems on January 14, 2012, and was obtained from the PVOL database (\href{http://pvol2.ehu.eus/}{http://pvol2.ehu.eus/}). }
     \label{fig:wind_map}
   \end{figure*}

The standing wave removal step applied to the SMA data using sine waves or polynomials, coupled to a lower spectral resolution compared to the ALMA data, results in too large uncertainties on the frequency scale. It thus makes any wind retrieval result highly suspicious and we decide not to apply the wind retrieval procedure to those data. On the other hand, and similarly to \citet{Carrion-Gonzalez2023}, the improved bandpass calibration, coupled with the continuum subtraction in the uv-plane, and a good signal-to-noise ratio on the CO lines at the planet limb, allow us to apply the wind retrieval algorithm of \citet{Cavalie2021} and \citet{Benmahi2022} and derive Doppler shifts from the ALMA data, even if the spectral resolution of 244\,kHz is lower than that of the data used in \citet{Cavalie2021} and \citet{Benmahi2022}. The LOS wind speeds we obtain are shown in \fig{fig:winds_summary} and provide coverage from $\sim$60$^\circ$S to $\sim$85$^\circ$N. The rings prevent meaningful measurements from 15$^\circ$S to 31$^\circ$S. The uncertainties in the derived values include the retrieval uncertainty (directly derived from the spectral noise) and other systematics discussed in detail in \citet{Cavalie2021}, especially $\sim$50 m/s caused by the spectral baseline removal. The zonal wind speeds, after correcting for the projection factor caused by the sub-Earth point latitude in January 2012 (15.1$^\circ$N) and the relative longitude with respect to the central meridian, are displayed in \figs{fig:wind_map}{fig:winds_diff} before and after averaging of the two limbs. The measured winds are located at the 0.2-1\,mbar level (see \fig{fig:contribution}), according to the wind contribution function calculation proposed by \citet{Lellouch2019}. These contribution functions are computed by weighting the monochromatic contribution functions at different relative frequencies from the line center by the local spectral slope of the line.

This data set provides the first measurements of stratospheric zonal winds in the southern hemisphere, as the data published in \citet{Benmahi2022,Benmahi2025} had been taken close to the northern summer solstice, i.e., when the southern hemisphere was either under the shadow of the rings or behind the limb. We can identify two prograde jets, centered around 50$^\circ$S and 40$^\circ$S, with speeds of 135$\pm$60\,m.s$^{-1}$ and 210$\pm$60\,m.s$^{-1}$. Although the first is relatively narrow with a width of about 5$^\circ$, the second extends over 10$^\circ$-15$^\circ$. Whether these are related to the tropospheric jets seen at 55$^\circ$S and 42$^\circ$S \citep{Garcia-Melendo2011} would require new ALMA observations with higher sensitivity and better spectral resolution to be taken when the southern hemisphere is fully observable, i.e, between the southern spring and autumn equinoxes.

\begin{figure*}[!t]
     \centering
     \includegraphics[width=0.95\textwidth]{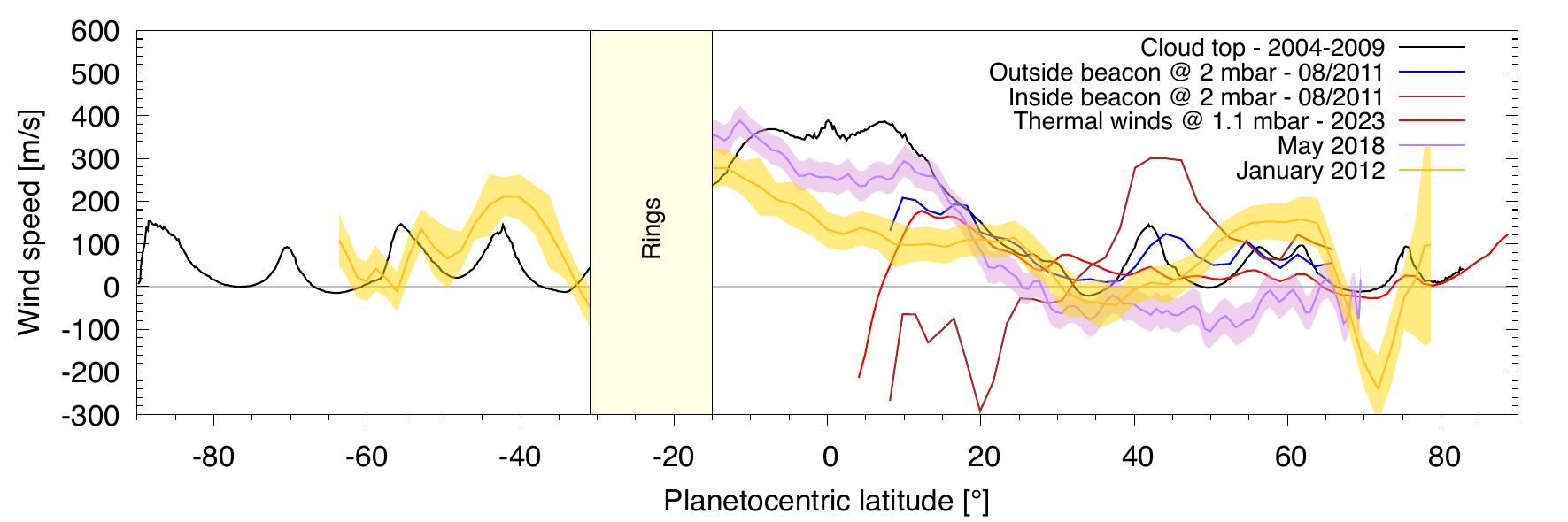}
     \caption{Comparison of average stratospheric wind speeds at 0.2-1\,mbar between January 2012 (gold line) and May 2018 (purple line) with their respective 1-$\sigma$ error bar envelopes. The January 2012 data are the average of the data shown in \fig{fig:winds_summary} after compensating for LOS projection. Error bars are also compensated for LOS projection. May 2018 data are taken from \citet{Benmahi2025}. For the purpose of comparison, cloud-top winds measured by \citet{Garcia-Melendo2011} with Cassini between 2004 and 2009 (black line), thermal winds on August 21st 2011 calculated at 2\,mbar by \citet{Fletcher2012b} outside beacon longitudes (blue line) and at the beacon central longitude (brown line), and thermal winds 1.1\,mbar from temperature retrievals applied to JWST observations of 2022 by \citet{Fletcher2023b} (red line), are overplotted.}
     \label{fig:winds_diff}
   \end{figure*}

The measurements confirm the presence of the broad prograde equatorial jet initially detected by \citet{Benmahi2022,Benmahi2025} in the stratosphere. The main difference lies in the weakening of the jet, with speeds reduced by about 100\,m.s$^{-1}$ in the southern hemisphere and up to 200\,m.s$^{-1}$ in the northern hemisphere. We also note that the jet appears broader in 2012 compared to 2018, extending up to $\sim$30$^\circ$N, whereas it was confined to $\sim$20$^\circ$N in 2018, i.e., close to northern summer solstice (see \fig{fig:winds_diff}). This confirms results obtained by \citet{Fletcher2017}, who found that the temperatures in the mbar layers in the equatorial region had been largely perturbed by the storm activity, inducing the injection of westward momentum in these layers according to the thermal wind balance, and thus leading a weakening of the jet in the mbar layers. We see no evidence for the signature of the Saturn quasiperiodic oscillation \citep{Fouchet2008,Guerlet2018}, most probably because of our too coarse spatial resolution. This lack of evidence may also result from the storm influence, as \citet{Fletcher2017} had observed a disruption of the oscillation from thermal data.

Around 40$^\circ$N, we tentatively detect the signature of the beacon on the wind pattern on the eastern limb. It is found in the form of a retrograde peak between 33$^\circ$N and 40$^\circ$N, and a prograde one between 40$^\circ$N and 47$^\circ$N, typical of anticyclonic motions. The peaks are located at latitudes where the temperature gradient is strong, i.e., between the central hot core and the background atmosphere, as seen in the thermal data of \citet{Fletcher2012b}. The peaks are symmetric in speed with values of 60$\pm$15\,m.s$^{-1}$. These speeds are weaker than those  indirectly derived at 2\,mbar from the thermal wind balance by \citet{Fletcher2012b} in August 2011 by a factor of 4-5 (see \fig{fig:winds_diff}). This is possibly due to four factors: (i) the beacon had already begun to weaken between the summer of 2011 and January 2012 \citep{Fletcher2012b}, (ii) the CIRS data lack information on the windshear (and therefore on the wind speeds) in the 5-70\,mbar range, making the integration of the winds with height prone to uncertainty, (iii) our limited spatial resolution of about 8$^\circ$ at the latitude of the beacon which smooths wind peaks, and (iv) the integration time of 1\,hr which smooths the velocities over 30$^\circ$ in longitude (the beacon longitudinal width was $\sim$50$^\circ$ ; see \fig{fig:temps2012}). On the opposite limb, the winds are tentatively found retrograde with an average speed of -50$\pm$20\,m.s$^{-1}$ between 30$^\circ$N and 45$^\circ$N. The beacon thus sat on a retrograde flow, thus explaining its observed retrograde drift rate \citep{Fletcher2012b,Lefour2025}. Such stratospheric retrograde motions at these latitudes had already been hinted by \citet{Benmahi2022,Benmahi2025} with speeds of -45$\pm$20\m.s$^{-1}$.

Beyond the latitude range perturbed by the beacon, our measurements are broadly different from those of \citep{Benmahi2022,Benmahi2025}. We identify two strong jets, one prograde and one retrograde, despite the reduced latitudinal resolution of the measurements. 

A broad 140$\pm$45\,m.s$^{-1}$ prograde jet, not seen in the thermal wind of \citet{Fletcher2012b} but evident as very strong windshear in Fig. 4 of \citet{Fletcher2018b}, extends from 50$^\circ$N to 67$^\circ$N. Whether this broad jet could be the stratospheric counterpart of the steady jet found at the cloud top \citep{Garcia-Melendo2011} is questionable, because it was not observed in the stratosphere in 2018 \citep{Benmahi2022,Benmahi2025}. This jet is presumably time-variable, maybe linked to seasonal evolution, because it was not detected by \citet{Benmahi2022,Benmahi2025}.

All jets found in these data are thus prograde, as is generally found in the troposphere \citep{Garcia-Melendo2011}, except a narrower retrograde jet that is located around 74$^\circ$N with a peak velocity of -220$\pm$80\,m.s$^{-1}$. This jet might be associated with the dissipating northern polar stratospheric vortex in 2010-2012, as these winds probe the right pressure level where radiative heating and possibly atmospheric subsidence was observed by \citet{Fletcher2018b}, also leading to negative windshear in this altitude and latitude region. This jet may alternately be caused by the auroral activity in the polar atmosphere of Saturn. The planet main UV emission is indeed produced between 70$^\circ$S and 80$^\circ$S \citep{Lamy2018}. Similar counterrotating polar jets, colocated with the main UV polar emission, have also been detected at 0.1\,mbar in Jupiter \citep{Cavalie2021}. However, Saturn has no tilt of its magnetic field axis with respect to its rotation axis, which would make any auroral jets zonal and thus more difficult to identify compared to Jupiter. The nature of the retrograde jet found around 74$^\circ$N thus remains elusive at this stage. If confirmed at Saturn, auroral jets may then be a generic feature resulting from auroral activity in giant planets with a strong magnetic field and intense auroral activity. We note that the complex magnetic field found at Uranus and Neptune may make it quite complex to verify if such aurorally induced jets exist in these planets.

\section{Conclusion}\label{sec:Conclusions}
In this paper, we have presented spatially and spectrally resolved observations of the stratospheric CO emission emanating from Saturn. The observations were taken in March 2010 with SMA and in January 2012 with ALMA, i.e., before and during the last Great Storm event. Regarding the distribution of CO, our main findings can be summarized as follow:
\begin{itemize}
    \item The CO data from the 2012 ALMA observations can be reproduced with fits of moderate quality with the rescaled 220-year-old-comet-impact profile from \citet{Cavalie2010}. The limited quality of the fits is most probably a result from the baseline ripple removal stage of the data reduction, which removed part of the broad wings of the CO line. It consequently limits our ability to obtain better fits of the broad wings of the line with the adopted comet profile.
    \item The mean CO mole fraction measured at 0.3\,mbar, i.e., at the peak of the contribution functions, is (1.7$\pm$0.7)\dix{-7}, according to the more accurate ALMA observations. Given the limitations in the data quality, we see no evidence for meridional variations.
    \item The SMA data suffer from larger error bars, with a mean CO mole fraction at 0.3\,mbar of (1.7$\pm$1.2)\dix{-7}, which do not enable us to identify measurable changes in the CO distribution, even at the location of the beacon.
\end{itemize}
Regarding stratospheric wind measurements, our results can be summarized as follow:
\begin{itemize}
    \item Despite limitations in latitudinal resolution, we tentatively detect the signature of the beacon in the form of an anticyclone sitting on a retrograde flow. This explains the retrograde drift rate of the beacon, as observed over the lifetime of the beacon \citep{Fletcher2012b}.
    \item We find that the broad equatorial jet is weaker in January 2012 than in May 2018 by about 100\,m.s$^{-1}$ in the southern part of the jet and by 200\,m.s$^{-1}$ from the equator to 20$^\circ$N. This observation confirms the slowing effect on the jet that resulted from the temperature perturbations caused by the storm at latitudes as low as the equatorial region that \citet{Fletcher2017} measured.
    \item We detect for the first time several prograde jets in the Saturn stratosphere, at 40$^\circ$S, 50$^\circ$S, and 60$^\circ$N, with speeds in the range of 100-200\,m.s$^{-1}$. Higher spatial resolution is required to confirm whether these jets are stratospheric counterparts of tropospheric jets measured since decades at the cloud-top.
    \item We detect a narrow retrograde jet centered at 74$^\circ$N with a peak velocity of -220$\pm$80\,m.s$^{-1}$. This jet may be tied to the main UV emission \citep{Lamy2018}, similarly to what was observed in Jupiter's polar stratosphere by \citet{Cavalie2021} or may result from the dissipating southern polar stratospheric vortex in 2010-2012 \citep{Fletcher2018b}.
\end{itemize}

Now that Saturn is nearing equinox again, new observations with higher spatial resolution are needed for several reasons. We first need to confirm the presence of the jets observed in the southern hemisphere, and determine whether a southern retrograde polar jet also exists. The observation of the equatorial jet will assess whether its speed remains consistent with the 2018 data or if it is slower during equinox. If the jet exhibits speeds similar to those observed in 2018, then the slower velocities measured during the storm would definitely be attributed to the perturbations induced by the storm. Finally, and even if the evaluation of the Saturn quasiperiodic oscillation was not possible given the limitation of the 2012 ALMA dataset, its temporal evolution still needs to be comprehensively evaluated.

\section*{Acknowledgements}
T. Cavali\'e and C. Lefour were supported by the Programme National de Plan\'etologie (PNP) of CNRS/INSU and by the Centre National d'\'Etudes Spatiales (CNES). Fletcher was supported by STFC Consolidated Grant reference ST/W00089X/1. T. Cavali\'e thanks Freddy Willems for enabling us to use his optical image of Saturn in this work.

This paper makes use of the following ALMA data: ADS/JAO.ALMA\#2011.0.00808.S. ALMA is a partnership of ESO (representing its member states), NSF (USA) and NINS (Japan), together with NRC (Canada), NSTC and ASIAA (Taiwan), and KASI (Republic of Korea), in cooperation with the Republic of Chile. The Joint ALMA Observatory is operated by ESO, AUI/NRAO and NAOJ. 

The Submillimeter Array is a joint project between the Smithsonian Astrophysical Observatory and the Academia Sinica Institute of Astronomy and Astrophysics and is funded by the Smithsonian Institution and the Academia Sinica.

We recognize that Maunakea is a culturally important site for the indigenous Hawaiian people; we are privileged to study the cosmos from its summit.

\bibliographystyle{aa} 

\begin{thebibliography}{47}
\expandafter\ifx\csname natexlab\endcsname\relax\def\natexlab#1{#1}\fi

\bibitem[{{Benmahi} {et~al.}(2022){Benmahi}, {Cavali{\'e}}, {Fouchet},
  {Moreno}, {Lellouch}, {Bardet}, {Guerlet}, {Hue}, \& {Spiga}}]{Benmahi2022}
{Benmahi}, B., {Cavali{\'e}}, T., {Fouchet}, T., {et~al.} 2022, \aap, 666, A117

\bibitem[{{Benmahi} {et~al.}(2025){Benmahi}, {Cavali{\'e}}, {Fouchet},
  {Moreno}, {Lellouch}, {Bardet}, {Guerlet}, {Hue}, \& {Spiga}}]{Benmahi2025}
{Benmahi}, B., {Cavali{\'e}}, T., {Fouchet}, T., {et~al.} 2025, \aap, 696, C5

\bibitem[{{Borysow} {et~al.}(1985){Borysow}, {Trafton}, {Frommhold}, \&
  {Birnbaum}}]{Borysow1985}
{Borysow}, J., {Trafton}, L., {Frommhold}, L., \& {Birnbaum}, G. 1985, \apj,
  296, 644

\bibitem[{{Borysow} \& {Frommhold}(1986)}]{Borysow1986}
{Borysow}, A. \& {Frommhold}, L. 1986, \apj, 304, 849

\bibitem[{{Borysow} {et~al.}(1988){Borysow}, {Frommhold}, \&
  {Birnbaum}}]{Borysow1988}
{Borysow}, J., {Frommhold}, L., \& {Birnbaum}, G. 1988, \apj, 326, 509

\bibitem[{{Butler}(2012)}]{Butler2012}
{Butler}, B. 2012, {ALMA Memo \#594 -- Flux Density Models for Solar System
  Bodies in CASA}, Tech. rep., NRAO

\bibitem[{{Carri{\'o}n-Gonz{\'a}lez} {et~al.}(2023){Carri{\'o}n-Gonz{\'a}lez},
  {Moreno}, {Lellouch}, {Cavali{\'e}}, {Guerlet}, {Milcareck}, {Spiga},
  {Cl{\'e}ment}, \& {Leconte}}]{Carrion-Gonzalez2023}
{Carri{\'o}n-Gonz{\'a}lez}, {\'O}., {Moreno}, R., {Lellouch}, E., {et~al.}
  2023, \aap, 674, L3

\bibitem[{{Cassidy} \& {Johnson}(2010)}]{Cassidy2010}
{Cassidy}, T.~A. \& {Johnson}, R.~E. 2010, \icarus, 209, 696

\bibitem[{{Cavali{\'e}} {et~al.}(2009){Cavali{\'e}}, {Billebaud}, {Dobrijevic},
  {Fouchet}, {Lellouch}, {Encrenaz}, {Brillet}, {Moriarty-Schieven},
  {Wouterloot}, \& {Hartogh}}]{Cavalie2009}
{Cavali{\'e}}, T., {Billebaud}, F., {Dobrijevic}, M., {et~al.} 2009, \icarus,
  203, 531

\bibitem[{{Cavali{\'e}} {et~al.}(2010){Cavali{\'e}}, {Hartogh}, {Billebaud},
  {Dobrijevic}, {Fouchet}, {Lellouch}, {Encrenaz}, {Brillet}, \&
  {Moriarty-Schieven}}]{Cavalie2010}
{Cavali{\'e}}, T., {Hartogh}, P., {Billebaud}, F., {et~al.} 2010, \aap, 510,
  A88

\bibitem[{{Cavali{\'e}} {et~al.}(2014){Cavali{\'e}}, {Moreno}, {Lellouch},
  {Hartogh}, {Venot}, {Orton}, {Jarchow}, {Encrenaz}, {Selsis}, {Hersant}, \&
  {Fletcher}}]{Cavalie2014}
{Cavali{\'e}}, T., {Moreno}, R., {Lellouch}, E., {et~al.} 2014, \aap, 562, A33

\bibitem[{{Cavali{\'e}} {et~al.}(2015){Cavali{\'e}}, {Dobrijevic}, {Fletcher},
  {Loison}, {Hickson}, {Hue}, \& {Hartogh}}]{Cavalie2015}
{Cavali{\'e}}, T., {Dobrijevic}, M., {Fletcher}, L.~N., {et~al.} 2015, \aap,
  580, A55

\bibitem[{{Cavali{\'e}} {et~al.}(2019){Cavali{\'e}}, {Hue}, {Hartogh},
  {Moreno}, {Lellouch}, {Feuchtgruber}, {Jarchow}, {Cassidy}, {Fletcher},
  {Billebaud}, {Dobrijevic}, {Rezac}, {Orton}, {Rengel}, {Fouchet}, \&
  {Guerlet}}]{Cavalie2019}
{Cavali{\'e}}, T., {Hue}, V., {Hartogh}, P., {et~al.} 2019, \aap, 630, A87

\bibitem[{{Cavali{\'e}} {et~al.}(2021){Cavali{\'e}}, {Benmahi}, {Hue},
  {Moreno}, {Lellouch}, {Fouchet}, {Hartogh}, {Rezac}, {Greathouse},
  {Gladstone}, {Sinclair}, {Dobrijevic}, {Billebaud}, \&
  {Jarchow}}]{Cavalie2021}
{Cavali{\'e}}, T., {Benmahi}, B., {Hue}, V., {et~al.} 2021, \aap, 647, L8

\bibitem[{{Cavali{\'e}} {et~al.}(2024){Cavali{\'e}}, {Lunine}, {Mousis}, \&
  {Hueso}}]{Cavalie2024}
{Cavali{\'e}}, T., {Lunine}, J., {Mousis}, O., \& {Hueso}, R. 2024, \ssr, 220,
  8

\bibitem[{{Dick} {et~al.}(2009){Dick}, {Drouin}, \& {Pearson}}]{Dick2009}
{Dick}, M.~J., {Drouin}, B.~J., \& {Pearson}, J.~C. 2009, \jqsrt, 110, 619

\bibitem[{{Fischer} {et~al.}(2011){Fischer}, {Kurth}, {Gurnett}, {Zarka},
  {Dyudina}, {Ingersoll}, {Ewald}, {Porco}, {Wesley}, {Go}, \&
  {Delcroix}}]{Fischer2011}
{Fischer}, G., {Kurth}, W.~S., {Gurnett}, D.~A., {et~al.} 2011, \nat, 475, 75

\bibitem[{{Fletcher} {et~al.}(2007){Fletcher}, {Irwin}, {Teanby}, {Orton},
  {Parrish}, {Calcutt}, {Bowles}, {de Kok}, {Howett}, \&
  {Taylor}}]{Fletcher2007}
{Fletcher}, L.~N., {Irwin}, P.~G.~J., {Teanby}, N.~A., {et~al.} 2007, \icarus,
  188, 72

\bibitem[{{Fletcher} {et~al.}(2011){Fletcher}, {Hesman}, {Irwin}, {Baines},
  {Momary}, {Sanchez-Lavega}, {Flasar}, {Read}, {Orton}, {Simon-Miller},
  {Hueso}, {Bjoraker}, {Mamoutkine}, {del Rio-Gaztelurrutia}, {Gomez},
  {Buratti}, {Clark}, {Nicholson}, \& {Sotin}}]{Fletcher2011}
{Fletcher}, L.~N., {Hesman}, B.~E., {Irwin}, P.~G.~J., {et~al.} 2011, \science,
  332, 1413

\bibitem[{{Fletcher} {et~al.}(2012){Fletcher}, {Hesman}, {Achterberg}, {Irwin},
  {Bjoraker}, {Gorius}, {Hurley}, {Sinclair}, {Orton}, {Legarreta},
  {Garc{\'\i}a-Melendo}, {S{\'a}nchez-Lavega}, {Read}, {Simon-Miller}, \&
  {Flasar}}]{Fletcher2012b}
{Fletcher}, L.~N., {Hesman}, B.~E., {Achterberg}, R.~K., {et~al.} 2012,
  \icarus, 221, 560

\bibitem[{{Fletcher} {et~al.}(2017){Fletcher}, {Guerlet}, {Orton}, {Cosentino},
  {Fouchet}, {Irwin}, {Li}, {Flasar}, {Gorius}, \&
  {Morales-Juber{\'{\i}}as}}]{Fletcher2017}
{Fletcher}, L.~N., {Guerlet}, S., {Orton}, G.~S., {et~al.} 2017, \natastron, 1,
  765

\bibitem[{{Fletcher} {et~al.}(2018){Fletcher}, {Orton}, {Sinclair}, {Guerlet},
  {Read}, {Antu{\~n}ano}, {Achterberg}, {Flasar}, {Irwin}, {Bjoraker},
  {Hurley}, {Hesman}, {Segura}, {Gorius}, {Mamoutkine}, \&
  {Calcutt}}]{Fletcher2018b}
{Fletcher}, L.~N., {Orton}, G.~S., {Sinclair}, J.~A., {et~al.} 2018,
  \natcommun, 9, 3564

\bibitem[{{Fletcher} {et~al.}(2023){Fletcher}, {King}, {Harkett}, {Hammel},
  {Roman}, {Melin}, {Hedman}, {Moses}, {Guerlet}, {Milam}, \&
  {Tiscareno}}]{Fletcher2023b}
{Fletcher}, L.~N., {King}, O. R.~T., {Harkett}, J., {et~al.} 2023, \jgr, 128,
  e2023JE007924

\bibitem[{{Fouchet} {et~al.}(2008){Fouchet}, {Guerlet}, {Strobel},
  {Simon-Miller}, {B{\'e}zard}, \& {Flasar}}]{Fouchet2008}
{Fouchet}, T., {Guerlet}, S., {Strobel}, D.~F., {et~al.} 2008, \nat, 453, 200

\bibitem[{{Fouchet} {et~al.}(2017){Fouchet}, {Lellouch}, {Cavali{\'e}}, \&
  {B{\'e}zard}}]{Fouchet2017}
{Fouchet}, T., {Lellouch}, E., {Cavali{\'e}}, T., \& {B{\'e}zard}, B. 2017, in
  AAS/Division for Planetary Sciences Meeting Abstracts, Vol.~49, AAS/Division
  for Planetary Sciences Meeting Abstracts \#49, 209.05

\bibitem[{{Garc{\'\i}a-Melendo} {et~al.}(2011){Garc{\'\i}a-Melendo}, {Arregi},
  {Rojas}, {Hueso}, {Barrado-Izagirre}, {G{\'o}mez-Forrellad},
  {P{\'e}rez-Hoyos}, {Sanz-Requena}, \&
  {S{\'a}nchez-Lavega}}]{Garcia-Melendo2011}
{Garc{\'\i}a-Melendo}, E., {Arregi}, J., {Rojas}, J.~F., {et~al.} 2011,
  \icarus, 211, 1242

\bibitem[{{Guerlet} {et~al.}(2011){Guerlet}, {Fouchet}, {B{\'e}zard}, {Flasar},
  \& {Simon-Miller}}]{Guerlet2011}
{Guerlet}, S., {Fouchet}, T., {B{\'e}zard}, B., {Flasar}, F.~M., \&
  {Simon-Miller}, A.~A. 2011, \grl, 38, L09201

\bibitem[{{Guerlet} {et~al.}(2018){Guerlet}, {Fouchet}, {Spiga}, {Flasar},
  {Fletcher}, {Hesman}, \& {Gorius}}]{Guerlet2018}
{Guerlet}, S., {Fouchet}, T., {Spiga}, A., {et~al.} 2018, \jgr, 123, 246

\bibitem[{{Hartogh} {et~al.}(2011){Hartogh}, {Lellouch}, {Moreno},
  {Bockel{\'e}e-Morvan}, {Biver}, {Cassidy}, {Rengel}, {Jarchow},
  {Cavali{\'e}}, {Crovisier}, {Helmich}, \& {Kidger}}]{Hartogh2011a}
{Hartogh}, P., {Lellouch}, E., {Moreno}, R., {et~al.} 2011, \aap, 532, L2

\bibitem[{{Hedman} \& {Showalter}(2016)}]{Hedman2016}
{Hedman}, M.~M. \& {Showalter}, M.~R. 2016, \icarus, 279, 155

\bibitem[{{Hesman} {et~al.}(2012){Hesman}, {Bjoraker}, {Sada}, {Achterberg},
  {Jennings}, {Romani}, {Lunsford}, {Fletcher}, {Boyle}, {Simon-Miller},
  {Nixon}, \& {Irwin}}]{Hesman2012}
{Hesman}, B.~E., {Bjoraker}, G.~L., {Sada}, P.~V., {et~al.} 2012, \apj, 760, 24

\bibitem[{{Lamy} {et~al.}(2018){Lamy}, {Prang{\'e}}, {Tao}, {Kim}, {Badman},
  {Zarka}, {Cecconi}, {Kurth}, {Pryor}, {Bunce}, \& {Radioti}}]{Lamy2018}
{Lamy}, L., {Prang{\'e}}, R., {Tao}, C., {et~al.} 2018, \grl, 45, 9353

\bibitem[{{Lefour} {et~al.}(2025){Lefour}, {Cavali{\'e}}, {Feuchtgruber},
  {Moreno}, {Fletcher}, {Fouchet}, {Lellouch}, \& {Hartogh}}]{Lefour2025}
{Lefour}, C., {Cavali{\'e}}, T., {Feuchtgruber}, H., {et~al.} 2025, \aap, 698,
  A66

\bibitem[{{Lellouch} {et~al.}(1995){Lellouch}, {Paubert}, {Moreno}, {Festou},
  {Bezard}, {Bockelee-Morvan}, {Colom}, {Crovisier}, {Encrenaz}, {Gautier},
  {Marten}, {Despois}, {Strobel}, \& {Sievers}}]{Lellouch1995}
{Lellouch}, E., {Paubert}, G., {Moreno}, R., {et~al.} 1995, \nat, 373, 592

\bibitem[{{Lellouch} {et~al.}(2005){Lellouch}, {Moreno}, \&
  {Paubert}}]{Lellouch2005}
{Lellouch}, E., {Moreno}, R., \& {Paubert}, G. 2005, \aap, 430, L37

\bibitem[{{Lellouch} {et~al.}(2019){Lellouch}, {Gurwell}, {Moreno}, {Vinatier},
  {Strobel}, {Moullet}, {Butler}, {Lara}, {Hidayat}, \&
  {Villard}}]{Lellouch2019}
{Lellouch}, E., {Gurwell}, M.~A., {Moreno}, R., {et~al.} 2019, \natastron, 3,
  614

\bibitem[{{Levy} {et~al.}(1993){Levy}, {Lacome}, \& {Tarrago}}]{Levy1993}
{Levy}, A., {Lacome}, N., \& {Tarrago}, G. 1993, \jms, 157, 172

\bibitem[{{Levy} {et~al.}(1994){Levy}, {Lacome}, \& {Tarrago}}]{Levy1994}
{Levy}, A., {Lacome}, N., \& {Tarrago}, G. 1994, \jms, 166, 20

\bibitem[{{Li} {et~al.}(2023){Li}, {de Pater}, {Moeckel}, {Sault}, {Butler},
  {deBoer}, \& {Zhang}}]{Li2023b}
{Li}, C., {de Pater}, I., {Moeckel}, C., {et~al.} 2023, \sciadv, 9, eadg9419

\bibitem[{{Moreno}(1998)}]{Moreno1998}
{Moreno}, R.~M. 1998, PhD thesis, Universit\'e Paris VI

\bibitem[{{Moses} {et~al.}(2015){Moses}, {Armstrong}, {Fletcher}, {Friedson},
  {Irwin}, {Sinclair}, \& {Hesman}}]{Moses2015}
{Moses}, J.~I., {Armstrong}, E.~S., {Fletcher}, L.~N., {et~al.} 2015, \icarus,
  261, 149

\bibitem[{Perry {et~al.}(2018)Perry, Waite~Jr., Mitchell, Miller, Cravens,
  Perryman, Moore, Yelle, Hsu, Hedman, Cuzzi, Strobel, Hamil, Glein, Paxton,
  Teolis, \& McNutt~Jr.}]{Perry2018}
Perry, M.~E., Waite~Jr., J.~H., Mitchell, D.~G., {et~al.} 2018, \grl, 45,
  10,093

\bibitem[{{S{\'a}nchez-Lavega} {et~al.}(2011){S{\'a}nchez-Lavega}, {del
  R{\'{\i}}o-Gaztelurrutia}, {Hueso}, {G{\'o}mez-Forrellad}, {Sanz-Requena},
  {Legarreta}, {Garc{\'{\i}}a-Melendo}, {Colas}, {Lecacheux}, {Fletcher},
  {Barrado y Navascu{\'e}s}, {Parker}, {International Outer Planet Watch Team},
  {Akutsu}, {Barry}, {Beltran}, {Buda}, {Combs}, {Carvalho}, {Casquinha},
  {Delcroix}, {Ghomizadeh}, {Go}, {Hotershall}, {Ikemura}, {Jolly}, {Kazemoto},
  {Kumamori}, {Lecompte}, {Maxson}, {Melillo}, {Milika}, {Morales}, {Peach},
  {Phillips}, {Poupeau}, {Sussenbach}, {Walker}, {Walker}, {Tranter}, {Wesley},
  {Wilson}, \& {Yunoki}}]{Sanchez-Lavega2011}
{S{\'a}nchez-Lavega}, A., {del R{\'{\i}}o-Gaztelurrutia}, T., {Hueso}, R.,
  {et~al.} 2011, \nat, 475, 71

\bibitem[{{S{\'a}nchez-Lavega} {et~al.}(2018){S{\'a}nchez-Lavega}, {Fischer},
  {Fletcher}, {Garcia-Melendo}, {Hesman}, {Perez-Hoyos}, {Sayanagi}, \&
  {Sromovsky}}]{Sanchez-Lavega2018}
{S{\'a}nchez-Lavega}, A., {Fischer}, G., {Fletcher}, L.~N., {et~al.} 2018, in
  Saturn in the 21st Century, ed. K.~H. {Baines}, F.~M. {Flasar}, N.~{Krupp},
  \& T.~{Stallard}, 377--416

\bibitem[{{Sayanagi} {et~al.}(2013){Sayanagi}, {Dyudina}, {Ewald}, {Fischer},
  {Ingersoll}, {Kurth}, {Muro}, {Porco}, \& {West}}]{Sayanagi2013}
{Sayanagi}, K.~M., {Dyudina}, U.~A., {Ewald}, S.~P., {et~al.} 2013, \icarus,
  223, 460

\bibitem[{{Sromovsky} {et~al.}(2013){Sromovsky}, {Baines}, \&
  {Fry}}]{Sromovsky2013}
{Sromovsky}, L.~A., {Baines}, K.~H., \& {Fry}, P.~M. 2013, \icarus, 226, 402

\bibitem[{{Waite} {et~al.}(2018){Waite}, {Perryman}, {Perry}, {Miller}, {Bell},
  {Glein}, {Grimes}, {Hedman}, {Cuzzi}, {Brockwell}, {Teolis}, {Moore},
  {Mitchell}, {Persoon}, {Kurth}, {Wahlund}, {Morooka}, {Hadid}, {Walker},
  {Nagy}, {Yelle}, {Ledvina}, {Johnson}, {Tseng}, {Tucker}, \&
  {Ip}}]{Waite2018}
{Waite}, J.~H., {Perryman}, R.~S., {Perry}, M.~E., {et~al.} 2018, \science,
  362, 51

\end{thebibliography}

\end{document}